\documentclass[fleqn,usenatbib]{mnras}

\usepackage{newtxtext,newtxmath}

\usepackage[T1]{fontenc}

\DeclareRobustCommand{\VAN}[3]{#2}
\let\VANthebibliography\thebibliography
\def\thebibliography{\DeclareRobustCommand{\VAN}[3]{##3}\VANthebibliography}

\usepackage{graphicx}	
\usepackage[usenames, dvipsnames]{color}
\usepackage{hyperref}
\usepackage[nolist]{acronym}
\usepackage{mathtools}
\usepackage{pifont}
\usepackage{pdflscape}
\usepackage[dvipsnames]{xcolor}
\usepackage[normalem]{ulem} 

\definecolor{linkcolour}{rgb}{0,0,1}

\hypersetup{colorlinks=true,allcolors=[rgb]{0,0,1}}
\newacro{icm}[ICM]{intracluster medium}
\newacro{agn}[AGN]{active galactic nucleus}
\newacro{smbh}[SMBH]{supermassive black hole}
\newacro{bcg}[BCG]{brightest cluster galaxy}
\newacro{cgm}[CGM]{circumgalactic medium}
\newacro{gmc}[GMC]{giant molecular cloud}

\newcommand{\ramses}[0]{{\sc ramses} } 
\newcommand{\tcc}[0]{$t_{\text{cc}}$}
\newcommand{\cloud}[1]{ #1$_\text{cl}$}
\newcommand{\wind}[1]{ #1$_\text{w}$}
\newcommand{\cmark}{\ding{51}}%
\newcommand{\xmark}{\ding{55}}%
\newcommand{\run}[1]{{\fontfamily{lmtt}\selectfont \textbf{#1}}}
\newcommand{\Tfloor}[0]{$T_\text{fl}$ }

\title[Cold Clouds in Galaxy Clusters]{Shattering and growth of cold clouds in galaxy clusters: the role of radiative cooling, magnetic fields and thermal conduction}

\author[]{
Fred Jennings,$^{1,2}$\thanks{E-mail: \href{mailto:Fred.Jennings@ed.ac.uk}{Fred.Jennings@ed.ac.uk}}
Ricarda S. Beckmann,$^{2}$
Debora Sijacki$^{2}$
and Yohan Dubois$^{3}$
\\
$^{1}$Institute for Astronomy, University of Edinburgh, Royal Observatory,  Blackford Hill, Edinburgh, EH9 3HJ,  United Kingdom\\
$^{2}$Institute of Astronomy and Kavli Institute for Cosmology, University of Cambridge, Madingley Rd, Cambridge CB3 0HA, United Kingdom\\
$^{3}$Institut d'Astrophysique de Paris, UMR 7095, CNRS, UPMC Univ. Paris VI, 98 bis boulevard Arago, 75014 Paris, France
}

\date{Accepted for Publication in MNRAS}

\pubyear{2021}

\begin{document}
\label{firstpage}
\pagerange{\pageref{firstpage}--\pageref{lastpage}}
\maketitle

\begin{abstract}
 In galaxy clusters, the hot intracluster medium (ICM) can develop a striking multi-phase structure around the brightest cluster galaxy. Much work has been done on understanding the origin of this central nebula, but less work has studied its eventual fate after the originally filamentary structure is broken into individual cold clumps. In this paper we perform a suite of 30 (magneto-)hydrodynamical simulations of kpc-scale cold clouds with typical parameters as found by galaxy cluster simulations, to understand whether clouds are mixed back into the hot ICM or can persist. We investigate the effects of radiative cooling, small-scale heating, magnetic fields, and (anisotropic) thermal conduction on the long-term evolution of clouds. We find that filament fragments cool on timescales shorter than the crushing timescale, fall out of pressure equilibrium with the hot medium, and shatter, forming smaller clumplets. These act as nucleation sites for further condensation, and mixing via Kelvin-Helmholtz instability, causing cold gas mass to double within 75 Myr. Cloud growth depends on density, as well as on local heating processes, which determine whether clouds undergo ablation- or shattering-driven evolution. Magnetic fields slow down but don't prevent cloud growth, with the evolution of both cold and warm phase sensitive to the field topology. Counter-intuitively, anisotropic thermal conduction increases the cold gas growth rate compared to non-conductive clouds, leading to larger amounts of warm phase as well. We conclude that dense clumps on scales of $500$~pc or more cannot be ignored when studying the long-term cooling flow evolution of galaxy clusters. 
\end{abstract}

\begin{keywords}
galaxies: clusters: intracluster medium
\end{keywords}



\section{Introduction} \label{sec:introduction}

Galaxy clusters are the largest virialized structures in the Universe. Most of their baryon content ($\ge 90\%$) can be found in the form of a hot \ac{icm} that fills the galaxy cluster and extends out to the virial radius \citep{Mcnamara_2007,Andrey_2012}. The remaining baryons are bound in a collection of mostly early-type galaxies with old stellar populations, with the most massive \ac{bcg} residing at their centre. The \ac{icm} is observed to possess a complex multi-phase structure, with hot X-ray emitting gas ($\geq10^8$~K) co-existing with cold, dense structures ($\lesssim 10^4$~K) which likely form via local thermal instabilities at radii of several tens of kpc from the cluster center \citep{McDonald_2011,McCourt2012, Yang2016, Beckmann2019, Hitesh_2021}. This cold gas is found in and around the \ac{bcg}, where it can be observed as visually spectacular H$\alpha$ emission nebulae which may fuel cold, clump accretion onto the \ac{bcg} and their central \ac{smbh} \citep{McDonald2010, Gaspari2013, Tremblay_2016, TremblayCombes2018}. The most well-known example of this is the H$\alpha$ filaments in the Perseus cluster \citep{Fabian1994, PetersonFabian2006, Fabian2008, McDonaldGaspari2018}, although filaments in many other clusters have now been observed \citep[e.g.][]{RusselMcDonald2017, Olivares2019, JimenezGallardo2021J, NorthDavis2021}. Cold molecular gas is also observed with radio observations of CO lines surrounding the \ac{bcg} at radii of $50$~kpc or so, as a result of inflows \citep{Edge_2001,Salom_2004, NorthDavis2021}, and most notably up to within 10 kpc of the central \ac{bcg} with ALMA \citep{Mcnamara2014, Russell2014, Fogarty2019}. This inflow is thought to increase the growth efficiency and the magnitude of the feedback from the \ac{smbh} \citep{Degraf2016}, with important implications for the morphology of the gas around the central object. 

The existence of this cold gas reservoir is closely related to one of the major open problems in cluster physics today is the \textit{cooling flow problem}. Na\"ive estimates of gas cooling in the \ac{icm} predict large scale cooling flows, but assume an idealized, homogeneous cooling flow \citep{Fabian1994}, assumptions that are quite untrue in the observed multi-phase \ac{icm}, which has a much more complex substructure. The cooling flow model also neglects any heating processes which can inject thermal energy into the cold gas, for example through feedback processes. When comparing the cooling rate of hot gas via direct X-ray observations to the star formation rates in clusters, the discrepancy between the two suggests that there must in fact exist at least one non-gravitational source of thermal energy which offsets some of the cooling in these flows \citep[e.g.][]{Fabian1994,Peterson2001, Peterson_2003, HudsonMittal2010}. This effect must be strong enough to suppress cooling rates by at least an order of magnitude. Understanding the lack of cold gas actually seen in clusters compared to estimates based on na\"ive cooling flow calculations is a crucial hurdle to be overcome by any complete model of galaxy cluster evolution. Much work on solving the cooling flow problem has been directed towards understanding this heating mechanism, and in particular towards understanding how energy injected by the central \ac{agn} powered by the \ac{smbh}  in the \ac{bcg} offsets radiative heating losses in the cluster over long periods of time \citep[see for example][]{SijackiSpringel2006,Vernaleo2006, McNamara2012, Gaspari2013, Gaspari2015, Talbot2021, Bourne&Sijacki&Ewald2021,Bourne&Sijacki2021}. \ac{agn} feedback mainly operates in two modes; \textit{kinetic}, which generates jets and bubbles; and a \textit{radiative/quasar} mode in luminous \ac{agn}, where powerful winds may be driven by magnetic or radiation pressure, potentially displacing the cold gas. Feedback within galaxy clusters has been observed directly, as evidenced by jet-inflated bubbles with correspond to the so-called ``X-ray cavities'', by e.g. \citet{Boehringer1993,Churazov2000,McNamara&Wise2000,Fabian&Sanders2000, Birzan2004, Randall2011,Liu&Sun2020, Ubertosi&Brunetti2021}.

One crucial component of the overall cooling cycle of galaxy clusters is the evolution of cold gas once it has condensed from the hot phase. Large-scale cluster simulations have concluded that extended cold structures tend to fragment into smaller clumps over time \citep{Gaspari2013,Li2017b,Beckmann2019}, but lack the resolution to follow the evolution of individual small, cold clouds and ultimately to predict their fate. There are a wide range of potential scenarios for their evolution, with different consequences for large-scale cluster cooling flows. On one extreme, clouds could be quickly destroyed and mixed back locally into the hot phase. On the other extreme, small clumps could act as nucleation sites for cold gas condensation, which would make clouds grow in mass over time and eventually deliver large quantities of cold gas to the central \ac{bcg}.  If they survive for long periods of time, cold clouds could form part of a "fountain" of cold gas in galaxy clusters; formed when the feedback from a central \ac{agn} breaks up inflowing filaments, they are accelerated in outflows and follow a ballistic trajectory only to slow down and eventually fall back towards the cluster centre, along with newly formed cold gas that has condensed via instabilities at radii of up to $50$ kpc \citep{Mcnamara_2007}. Clouds that reach the \ac{bcg} may be then accreted by its central \ac{smbh} and ultimately help to determine the behaviour of the very feedback that initially created them by increasing gas accretion rates and reorientation the spin axis of the \ac{agn} jet. Furthermore, cold clumps should be sufficiently destroyed by jets and/or mixed through further interaction with the \ac{icm} if feedback models are to successfully solve the cooling flow problem and explain the lack of cold gas in these systems. The simulations of \citet{Beckmann2019} show that a large fraction of cold gas originally in filaments can survive on small scales after interaction with the jets. Therefore understanding how cold clumps might survive or perish in the hot gas within an explicit cluster-centric parameter space is key to the building of consistent models of both cluster cooling and heating, and of the coupling of feedback to the \ac{icm}. 

Studying individual clouds embedded in a hot background medium that is moving at a relative velocity to the cloud is known as the \textit{cloud-crushing problem} based on work by \cite{Klein1994} who calculated the analytic timescale on which small clumps get destroyed in such a scenario. Early simulations confirmed analytic expectations that clouds get disrupted and mix into the hot background medium \citep{Klein1994,Cooper_2009,Zhang2017,Scannapieco_2015}.  More recently, several authors have shown that in the presence of radiative cooling cold clouds can grow in some regions of the \ac{cgm} parameter space \citep{armillotta_fraternali_2016,McCourt_2017,Sparre2019,Gronke_Peng_2020,Kanjilal2021} due to efficient mixing, but get still get destroyed if they are too small. At the surface between the hot and cold gas, a mixed warm phase forms due to the Kelvin-Helmholtz instability \citep{Fielding2020}. When the cooling times of this warm gas are sufficiently short in comparison to the cloud crushing time, the cloud can grow over time (see Section \ref{sec:cloud_crushing_problem} for details). Other non-thermal physics, such as magnetic fields and thermal conduction \citep{armillotta_fraternali_2016,Bruggen2016,Li2020,SparrePfrommer2020} can also influence the long-term mass evolution of cold clouds. 

However, most of the work in the field has been conducted for a parameter space appropriate for the \ac{cgm} and associated cold clouds, rather than the remnants of H$\alpha$ filaments embedded in the \ac{icm} of a massive galaxy cluster. In this paper, we build on work in \citet{Beckmann2019}, who demonstrated how cooling filaments in simulations are broken up into small (100 pc-scale) clumps by interactions with the bimodal jets produced by a central \ac{smbh}. Being limited at a minimum resolution of $120 \rm \ pc$, this cluster-scale work was unable to follow the evolution of clouds on smaller scales. In the paper presented here, we re-simulate individual clouds with parameters typical to those seen in \citet{Beckmann2019} and the magnetised version of the same cluster \citet{Beckmann2022,Beckmann2022b} to understand the long-term evolution of such dense gas fragments, including the impact of radiative cooling, small-scale cloud reheating, magnetic fields and (anisotropic) thermal conduction.

The paper is structured as follows; in Section~\ref{sec:cloud_crushing_problem} we review the theory behind the cloud crushing problem and present the setup of our simulations in Section~\ref{sec:simulation_overview}. The non-magnetised cloud-crushing problem is investigated in Section~\ref{sec:cooling}, with the impact of magnetic fields studied in Section~\ref{sec:mhd} and that of thermal conduction in Section~\ref{sec:thermal_conduction}. Discussion and our conclusions are presented in Section~\ref{sec:conclusions}.

\section{The Cloud-Crushing Problem}
\label{sec:cloud_crushing_problem}

The classic set-up usually studied in the field is the \textit{cloud-crushing problem}. This is an idealised set-up that studies how an individual cloud, embedded in a (usually) uniform background wind, evolves over time. The characteristic timescale for the shredding of the cloud of uniform gas density $\rho_{\rm cl}$ and size $r_{\rm cl}$ within a uniform wind density $\rho_{\rm wind}$ and relative velocity of $v_{\rm wind}$ is the \textit{cloud-crushing timescale} \citep{Klein1994},
\begin{equation}\label{formula:tcc}
    t_\text{cc} \equiv \sqrt{\frac{\rho_{\text{cl}}}{\rho_\text{wind}}} \frac{r_\text{cl}}{v_\text{wind}} = \sqrt{\chi} \ t_\text{cr}\,,
\end{equation}
over which Kelvin-Helmholtz instabilities and shocks will cross the cloud radius and shred the cloud material. The non-magnetised, non-radiative cloud crushing problem is  \textit{scale-free} and can be entirely parameterised by the dimensionless density contrast $\chi$ and the cloud-crossing time $t_\text{cr}$. The actual physical size, wind velocity, and cloud/wind densities can then be re-scaled once these two parameters are determined. However, this approach breaks down in environments where radiative cooling in and around the cloud becomes important. This happens when the cooling time, 
\begin{equation} \label{formula: tcool}
t_{\text{cool}} \equiv \frac{\rho \varepsilon}{\mathcal{L}} = \frac{3 n k_{\rm B} T}{2 n_{\rm H}^2 \Lambda(T,Z)} \,,
\end{equation}
is comparable to or shorter than the cloud-crushing time. Here $\Lambda(T,Z) \equiv \mathcal{L} /n_{\rm H}^2$ is the \textit{cooling function}, $k_{\rm B}$ is the Boltzmann constant, $\varepsilon$ the specific energy, $T$ the temperature, $n$ the number density, and $n_{\rm H}$ is the hydrogen number density \citep[][Chapter 8]{Mo_White_Bosch}.

The ratio of cooling to cloud-crushing timescales for clouds is
 \begin{equation} \label{eqn: ratio of timescales}
\theta \equiv \frac{t_\text{cool}}{t_\text{cc}} = \frac{3 n_\text{cl} k_{\rm B} T_\text{cl}v_\text{wind}}{2 n^2_\text{H,cl} \Lambda (T,Z) r_\text{cl} \sqrt{\chi}} \,.
\end{equation}
The cloud-crushing is sub-dominant when the cloud is at much higher density than the wind (large $\chi$), and when the cooling is efficient (i.e. $\Lambda(T,Z)$ is large); for the metallicity of $Z=0.33 Z_\odot$ that we use (for both cloud and wind), the cooling function is strongly peaked between $10^5 - 10^6$ K \citep{Sutherland1993}, meaning that the radiative loses should not be ignored, unlike studies of largely pristine environments such as the \ac{cgm} where cooling may not be so important. This metallicity that we use is a well-known typical value for observations of the \ac{icm}, at least locally \citep[see for example][]{Mushotzky1978,Allen&Fabian1998,Tozzi2003,DeGrandi2004,Leccardi2008, Werner2013, Simionescu2015,Urban2017}.

\begin{figure}
	\includegraphics[width=\columnwidth ]{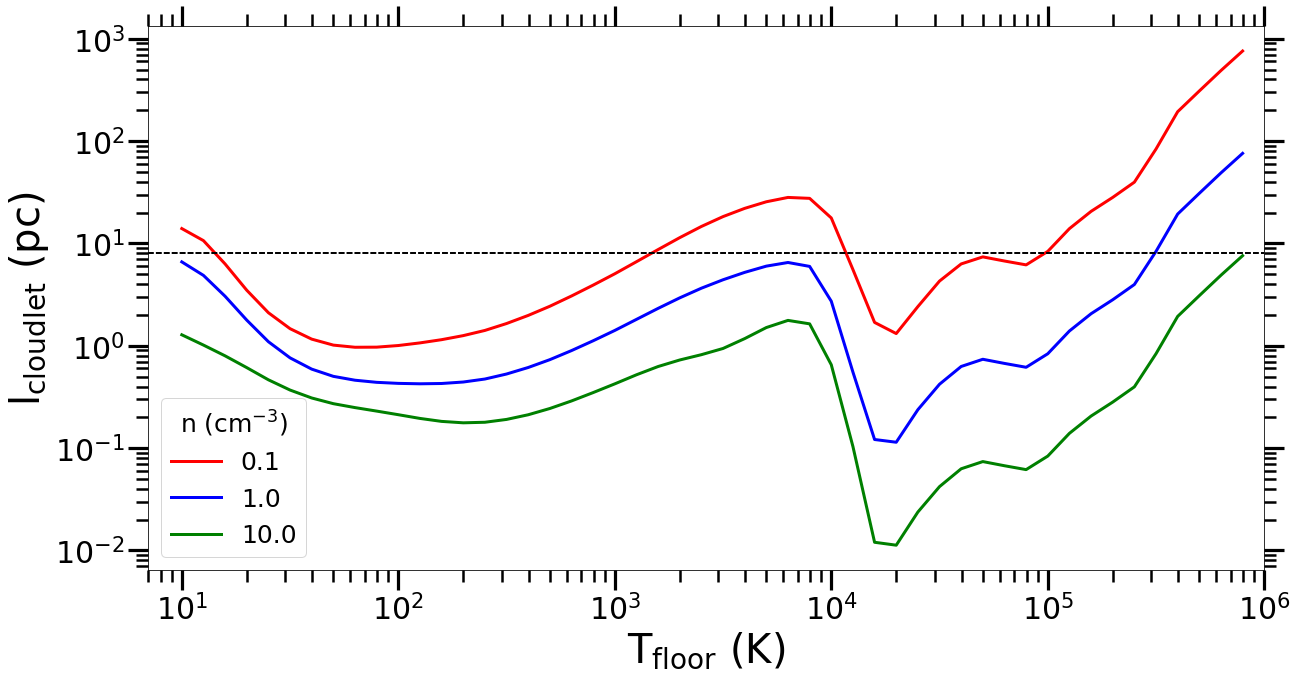}
    \caption{The minimum expected cloud size according to Equation~(\ref{eq:shattering_lengthscale}) as a function of the cold gas temperature for three different cloud number densities $n$. This plot assumes a metallicity of $Z=0.33 Z_\odot$, with solar abundances of elements, and an adiabatic index of $\gamma=5/3$. In general, denser clouds fragment to smaller clumps, but the evolution with temperature is not linear. The black horizontal line denotes our maximum spatial resolution of $8$~pc adopted for our main production runs, but see Appendix~\ref{appendix: convergence} for higher resolution simulation results.}
    \label{plot:shattering lengthscale}
\end{figure}

We can also equate the cooling and cloud-crossing timescales to obtain a characteristic length at which the cloud should be able to respond dynamically to cooling-induced perturbations in the temperature and pressure \citep{McCourt_2017}. This length $l_\text{cloudlet} \sim c_{\rm s}t_\text{cool}$ can be straightforwardly shown, for a fully ionised plasma with hydrogen fraction $X$ (assuming an ideal gas speed of sound), to be equal to
\begin{equation} \label{eq:shattering_lengthscale}
    l_\text{cloudlet} = \frac{3}{64} \left[\frac{ \gamma k_{\rm B}^3 T_\text{fl}^3}{m_{\rm p} X^4 n^2 \Lambda^2(T_\text{fl},Z)}(3 + 5X -Z)^5  \right]^{1/2} \,,
\end{equation}
where $\gamma$ is the adiabatic index of the gas and $T_{\rm fl}$ is the temperature floor that the gas settles at, and $m_p$ is the proton mass. The length-scale on which clumps are dynamically responsive is inversely proportional to the number density and the value of the cooling function, and proportional to the temperature raised to the $3/2$ power. We plot this length scale as a function of temperature in Fig.~\ref{plot:shattering lengthscale} for reference, using the cooling tables for dust-free, no cosmic rays, and no self-shielding model of \citet{PloeckingerSchaye2020}, who have generated cooling rates using the spectral synthesis code \href{https://gitlab.nublado.org/cloudy/cloudy/-/wikis/home}{{\sc cloudy}} \citep{Ferland2017}. Fig.~\ref{plot:shattering lengthscale} shows the analytical solution for the minimum theoretical lengthscale (given by Equation~(\ref{eq:shattering_lengthscale})) that is expected in cloud shattering, if the cloud shatters continuously down to a floor temperature (given on the horizontal axis), and thus gives the rough size of the smallest clumps which should be seen in simulations, resolution-permitting. Generally $l_{\rm cloudlet} < 1 \rm \ pc$ but very diffuse clouds in the temperature range $10^3-10^4 \rm \ K$ can have minimum sizes of $5 \rm \ pc$ or more. The minimum cell length on our fine grid in the simulations presented in this paper is $8$~pc for the majority of the runs, and we test a variety of temperature floor temperatures.

From the analytical result we see that we can expect to be able to only resolve clumps within a narrow range, that are around $0.5$ orders of magnitude larger than our resolution lengthscale, although we get close to resolving the length floor over a broad range for diffuse clouds. We would expect further shattering beyond most of the smallest clumps produced in our simulation runs, and to gauge what effect this may have we perform a convergence test in Appendix~\ref{appendix: convergence} using a range of resolutions that are computationally feasible. We find that with increased resolution less cold gas forms initially (by up to $40\%$ for t < \tcc), but at later times more cold gas forms due to the further shattering and higher mixing surface area-to-volume ratio. Therefore, we would expect that the our findings on the cold phase growth based on lower resolution simulation represent a conservative lower bound on the magnitude of this effect

\section{Numerical method} \label{sec:simulation_overview}

\subsection{Nomenclature}
Throughout this paper, we define a \textit{cold phase} to include any gas at a temperature lower than the initial cloud temperature $T<T_{\rm cl}$, a \textit{warm phase} to include gas between the initial cloud and wind temperatures ($T_{\rm cl} < T < T_{\rm wind}$), and a \textit{hot phase} to include any gas at or above the initial wind temperature $T \geq T_{\rm wind}$. For all quantitative measures of the three phases we see in the simulations, we use a straightforward temperature cut on the gas. For all simulations presented here, $T_{\rm cl}= 10^6 \rm \ $ and $T_{\rm wind} = 10^8 \rm \ K$. This wind/\ac{icm} temperature is consistent with X-ray observations of clusters \citep{Reiprich2002,HudsonMittal2010,Frank&Peterson2013} which measure X-ray temperatures mainly in the range $1-10$ keV.   The cloud temperature we use is consistent with the maximum temperature of so-called \textit{dense gas} used to identify the clump objects generated within the cluster in \citet{Beckmann2019}.

We also define as $T_{\rm fl}$ the minimum temperature to which gas is allowed to cool via radiative loses, the so-called cooling floor temperature.
Furthermore we note another important temperature measure, the mixing temperature $T_{\rm mix}\simeq\sqrt{T_{\rm cl}T_{\rm wind}}$ given by \cite{BegelmanFabian1990}, which is convenient for measuring some length scales of the problem (see \citealp{Gronke_Peng_2020}).

\subsection{General setup} \label{sec:ramses}
The simulations presented in this paper were produced using the {\sc ramses} code \citep{Teyssier2002}, an adaptive-mesh refinement magneto-hydrodynamic (MHD) code. The simulations are conducted in 3D and include radiative cooling for all of our production runs, apart from a numerical experiment where we test the non-radiative setup to highlight the differences. A subset of simulations also includes magnetic fields with different strengths and orientation, and (anisotropic) thermal conduction. To solve the MHD equations, \ramses employs a second-order Godunov method \citep[see][]{Toro2009}, with a MUSCL-Hancock scheme to solve the Euler equations and compute fluxes across cells. \ramses uses an octree refinement method to adaptively refine the space, and models magnetic fields as face-centred quantities with constrained transport to satisfy the solenoidality constraint $\mathbf{\nabla \cdot B = 0}$~\citep{Fromang2006}. A HLLC Riemann solver~\citep{Toro2009} was used for simulations without magnetic fields, with a HLLD solver~\citep{Miyoshi_2008} used for later MHD runs. The Courant factor was set to 0.8 for adaptive time-stepping, for which \ramses uses a second-order midpoint scheme. We use the {\sc yt Project} Python package \citep{Turk_2010} for all analysis.

\subsection{Initial conditions and refinement}
\label{sec:initial_conditions}
All simulations presented in this paper have a similar setup: a spherical cold cloud of radius $r_{\rm cloud} = 500$~pc and temperature $T_{\rm cl}= 10^6 \rm \ K$ is placed in a uniform hot background medium of temperature $T_{\rm wind}=10^8 \rm \ K$. We fix the initial cloud density contrast to $\chi=\rho_{\rm cl}/\rho_{\rm w} = 100$, but vary both the initial cloud density $\rho_{\rm cl}$ and the initial wind density $\rho_{\rm w}$ (see Table~\ref{tab:hydro runs} for details). This cloud temperature was selected in order to initialise the clouds in pressure equilibrium with the surrounding hot medium, whose temperature is typical for the \ac{icm} of galaxy clusters. This overdensity was chosen as the typical overdensities in the simulations of \citet{Beckmann2019} range from around 2-4 (see their Fig. 4). We select the lowest value in this range because this corresponds to the smallest crushing timescale, meaning we can simulate longer into the dynamical lifetime of the cloud.

Simulations are conducted in the frame of the cold cloud, which is initialised with zero bulk velocity. Instead, the background medium is given a bulk velocity of $200$ km\,s$^{-1}$, in line with observations of line-of-sight velocities in cluster centres \citep{Salome&Combes2006,McDonald&Veilleux2012,Tremblay_2016, Olivares2019}, and corresponding to the peak of the velocity distribution of the clumps produced in \citet{Beckmann2019} (see Fig. 7 of that work). This is equivalent to a cloud Mach number of $\mathcal{M}=2$. A random velocity dispersion field of normal distribution centred around zero and with standard deviation of $50$ km\,s$^{-1}$ within the hot medium and $30$ km\,s$^{-1}$ within the cloud was added to the cloud and wind bulk velocities in order to avoid grid-locking effects and roughly model the turbulence which may play an important role in galaxy clusters  \citep[e.g.][]{Churazov2012,GaspariChurazov2013, Schuecker2014,Walker2015, Hofmann2016, Zuhone2018, HitomiCollab2018}. We calibrated the mean and variance of the velocity dispersion to values found in and around clumps produced by the simulations presented in \citet{Beckmann2019}. No further turbulence driving is added throughout the simulation.  

The simulation is initialised with a passive scalar, initially set to unity within the cloud and zero outside, which is advected with the flow throughout the simulation. This scalar is used both during the simulation run and in post-processing to help distinguish mixed material from pristine background gas, as cells with high $f_{\rm scalar} = \rho_{\rm scalar}/\rho_{\rm gas}$ are dominated by gas originally found in the cloud, while those with low $f_{\rm scalar}$ are dominated by gas initially found in the hot background medium. Note however that quantitative measures of masses for the three fluid phases that we will discuss later on, use only a cut on the temperature of the plasma, as discussed above.

At the start of our simulations the cloud is setup to be in pressure equilibrium with the \ac{icm} in all runs, as this is appropriate in a galaxy cluster environment. All simulations were initiated with a metallicity of $Z=0.33 Z_\odot$, as is typical for clusters \citep{Baumgartner2005, Lovisari2019}, and this metallicity was kept constant throughout. The metallicity of the cloud is not expected to deviate significantly from the background medium, since clouds originally likely condense from the hot \ac{icm}. In practice, a metallicity gradient within the cluster and star formation occurring in and around cold clumps may lead to a range of different metallicity values for dense clumps, but note that the observed metallicity gradients of the \ac{icm} are typically shallow \citep{Leccardi2008}.

No gravity is included in our simulations. The free-fall time of our clouds is $t_\text{ff} = \sqrt{3\pi/32 G \bar{\rho_\text{cl}}}$ which, for our fiducial run (the lowest density we test), means the free-fall time is approximately $150$~Myr which corresponds to around $6 \ t_\text{cc}$. As $t_{\rm cool} < t_{\rm cc}$ in all of our radiative setups, there is a pressure-driven collapse on sub-crushing timescales during which gravity is sub-dominant and therefore not included explicitly here.

In MHD simulations, the magnetic field was intialised as a uniform field that is either initially parallel (``aligned'') or perpendicular (``normal'') to the flow. Thermal conduction is modelled following \cite{Dubois2016}, with a thermal conduction coefficient equal to $\kappa_{\rm Spitzer} = n_{\rm e} k_{\rm B} D_{\rm C}$ \citep{Spitzer1962}, where $n_{\rm e}$ is the electron number density, $k_{\rm B}$ is the Boltzmann constant and $D_{\rm C} = 8 \times 10^{30} \big(\frac{k_{\rm B} T_{\rm e}}{10^8 \rm \ K}\big)^{\frac{5}{2}} \big(\frac{n_{\rm e}}{10^{-2} \rm \ cm^{-3}}\big)^{-1} \rm \ cm^2 s^{-1}$ the thermal diffusivity and $T_{\rm e}$ the electron temperature. In the case of anisotropic thermal conduction, the perpendicular conductivity is set to $\kappa_{\rm perp} = 0.0001 \kappa_{\rm Spitzer}$.

The size of the simulation box is chosen such that the flow crossing time at the wind velocity is several tens of crushing times. In practice we never simulate for this long, and all runs are stopped before the mixed tail of the cloud reaches the edge of the simulation domain, so no material is lost from the simulation. Most simulations were performed in a box size of $65$ kpc, with a root grid of $32^3$ (refinement level 5) cells which was adaptively refined by 8 more levels up to a maximum resolution of $\Delta x = 8$ pc. The initial conditions are refined using concentric spheres centered on the cloud, with the highest refinement level incorporating the entire initial cloud. Refinement then proceeds adaptively where a cell is refined at the highest level of refinement if $f_{\rm scalar} > 10^{-6}$, and progressively derefined otherwise. 

All simulation parameters are summarised in Table~\ref{tab:hydro runs} for the hydrodynamical runs and in Table~\ref{tab:mhd runs} for MHD simulations with and without conduction.

\begin{table*} 
	\centering
	\caption{The parameters of the hydrodynamical runs, with the columns showing: (1) the run labels, (2) whether the run includes radiative cooling, (3) the initial cloud density $\rho_{\mathrm{cl}}$, (4) the initial wind (background ICM) density $\rho_{\mathrm{w}}$, (5) the length of the simulation box $d_\text{box}$, (6) the size of the highest resolution element $\Delta x$, (7) the value of the minimum temperature achieved through cooling $\rm T_{\rm fl}$, (8) the initial mass of the cold cloud. The crushing timescale \tcc\ for all runs is $24.4$~Myr, and the background wind velocity is $200$~km\,s$^{-1}$ in all cases. All simulations have an initial background temperature of $T_{\rm wind} = 10^8$~K, and an initial cloud temperature of $\rm T_{\rm cl} = 10^6$~K,except for \run{HD0.1\_1e4T0}, which was initialised with $\rm T_{\rm cl} = 10^4$~K}.
	\label{tab:hydro runs}
	\begin{tabular}{lccccccccr} 
		\hline
    	Label &  Cooling &  \cloud{$\rho$} [H/cm$^{3}$] & \wind{$\rho$} [H/cm$^{3}$]  &  $d_{\text{box}}$ [kpc] & $\Delta x$ [pc] & $\rm T_{\rm fl}$ [K]& $\rm M_{\rm 0}$ [$\rm M_\odot$]\\
    	\hline
        \run{HD10\_nc} &\xmark    & $10$ & $0.1$    & 150 & 18 & N/A & $1.28 \times 10^8$\\
        \run{HD0.1\_nc} & \xmark   & $0.1$ & $0.001$    & 150 & 18 & N/A & $1.28 \times 10^6$\\
        \run{HD10} &\cmark   & $10$ & $0.1$    & 65 &  8 & 10 & $1.28 \times 10^8$\\
        \run{HD1.0} &\cmark   & $1$ & $0.01$  & 65 & 8 & 10 & $1.28 \times 10^7$ \\
        \run{HD0.1} &\cmark   & $0.1$ & $0.001$   & 65 & 8 & 10 & $1.28 \times 10^6$\\
        \run{HD0.1\_1e4T0} &\cmark   & $0.1$ & $0.001$   & 65 & 8 & 10 & $1.28 \times 10^6$\\
        \run{HD0.1\_$F$5.5} &\cmark   & $0.1$ & $0.001$    & 65 &  8& $3 \times 10^5$ & $1.28 \times 10^6$\\
        \run{HD0.1\_$F$5.0} &\cmark   & $0.1$ & $0.001$    & 65 &  8& $1 \times 10^5$ & $1.28 \times 10^6$\\
        \run{HD0.1\_$F$4.5} &\cmark   & $0.1$ & $0.001$    & 65 &  8& $3 \times 10^4$ & $1.28 \times 10^6$\\
        \run{HD0.1\_$F$4.0} &\cmark   & $0.1$ & $0.001$    & 65 &  8& $1 \times 10^4$ & $1.28 \times 10^6$\\
        \run{HD0.1\_$F$3.5} &\cmark   & $0.1$ & $0.001$    & 65 &  8& $3 \times 10^3$ & $1.28 \times 10^6$\\
        \run{HD0.1\_$F$3.0} &\cmark   & $0.1$ & $0.001$    & 65 &  8& $1 \times 10^3$ & $1.28 \times 10^6$\\
    	\hline
    	\end{tabular}
\end{table*}

\subsection{Cooling}
Radiative cooling in \ramses is computed using values given in the cooling tables of \citet{Sutherland1993} above $10^4\,\rm K$ and those from~\cite{Rosen_alexander_1995} for temperatures below. In our simulations, gas cools to a minimum fiducial temperature of $T_{\rm fl} = 10 \rm \ K$ (although we do run a suite of tests investigating the effect of changing this value in Section~\ref{sec:Cooling Floor}).

In galaxy clusters, both cloud dynamical and cloud crushing timescales for cold clouds are comparable to or longer than the cooling time of the hot \ac{icm}. However, observationally, the \ac{icm} temperature does not change significantly over sufficiently long timescales, as the bulk of cooling of the \ac{icm} in galaxy clusters is offset by various heating sources, from energy injected by the \ac{agn} to possibly thermal conduction. To mimic the impact of such background heating in the simulations presented here, radiative cooling is only permitted in gas above a minimum concentration of a passive scalar originally only placed in the cold cloud (see Section~\ref{sec:initial_conditions}). This approach prevents the catastrophic cooling of the hot \ac{icm} background gas (as opposed to clump gas and mixed gas, which we allow to cool freely) which would otherwise occur over timescales which are short compared to the dynamical timescale, at variance with observations. We set the required passive scalar value to $f_{\rm scalar  } > 0.001 $, and still permit hot gas to cool via in-cell mixing with colder fluid. In MHD simulations, spurious magnetic heating occurred in a very small number of cells in and around the clouds. To prevent such ``hotspots'' from occurring as well as from slowing the simulation, we enforced a minimum density floor of $0.0005 \rm \ cm^{-3}$. Convergence tests showed no statistically significant impact on the long-term evolution of simulations from such a density floor.

\subsection{Clump finding}
\label{sec:clump_finding}

To gather statistics on the distribution of individual clumps in our simulations, we use the water-shed segmentation-based structure finding algorithm PHEW \citep{Bleuler2015}. Clumps are identified in all gas with $T<3 \times 10^{7} \rm \ K$, and densities above $n > 0.02$ cm$^{-3}$. All clumps with a relevance (peak-to-saddle ratio) of less than 30 are merged through the saddle into the neighbouring clump.

\section{Results} 
\label{sec:results}

\subsection{Radiative Cooling and Cloud Growth}
\label{sec:cooling}
In this section we study the impact of varying cloud and background gas parameters to understand how local gas properties impact the evolution of cold clumps in galaxy clusters. All simulations presented here have no magnetic fields, while the simulations with magnetic fields are discussed in Section~\ref{sec:mhd}. Specific simulation parameters are outlined in Table~\ref{tab:hydro runs}.

\subsubsection{The impact of radiative cooling}
 Two runs were performed with cooling turned off in order to quantify the impact of cooling and compare our work against the large existing body of work that does not consider cooling. 

\begin{figure}
	\includegraphics[width=\columnwidth ]{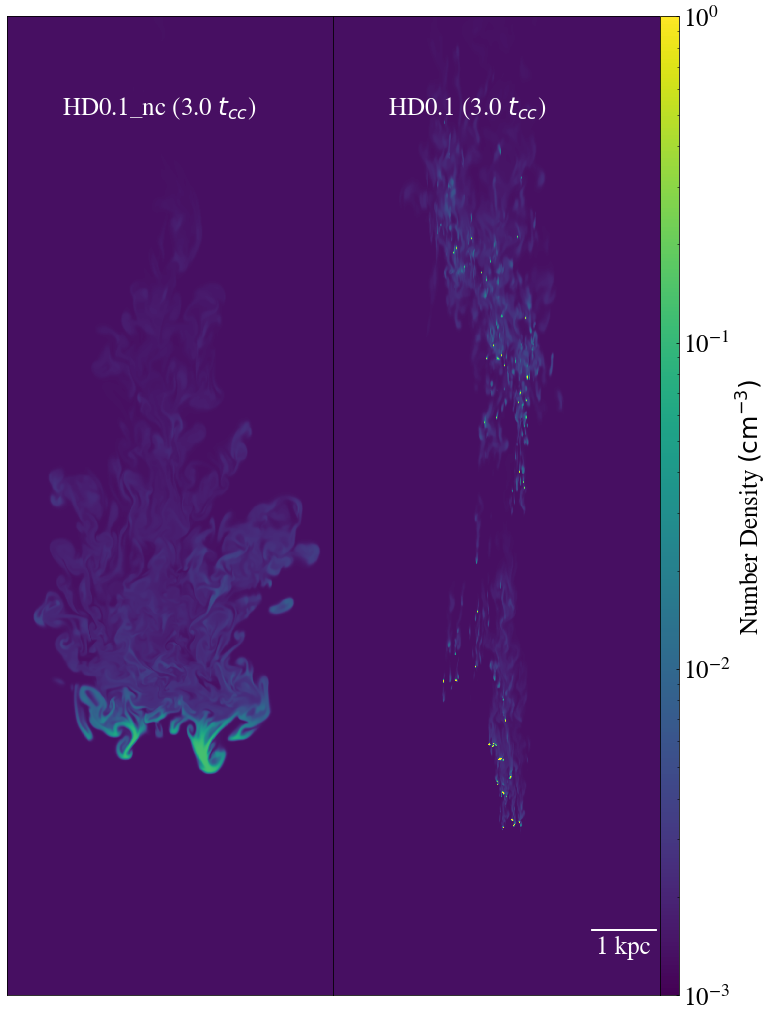}
    \caption{Number density slices showing the difference in evolution after 3 cloud-crushing times between the run without cooling (\run{HD0.1\_nc}, left) and an identical cloud run with cooling (\run{HD0.1}, right). Radiative cooling promotes shattering of the cloud during the early evolution, with cloud morhology being distinctly different.}
    \label{plot:HD0.1_nc vs HD0.1 density}
\end{figure}

\begin{figure}
	\includegraphics[width=\columnwidth ]{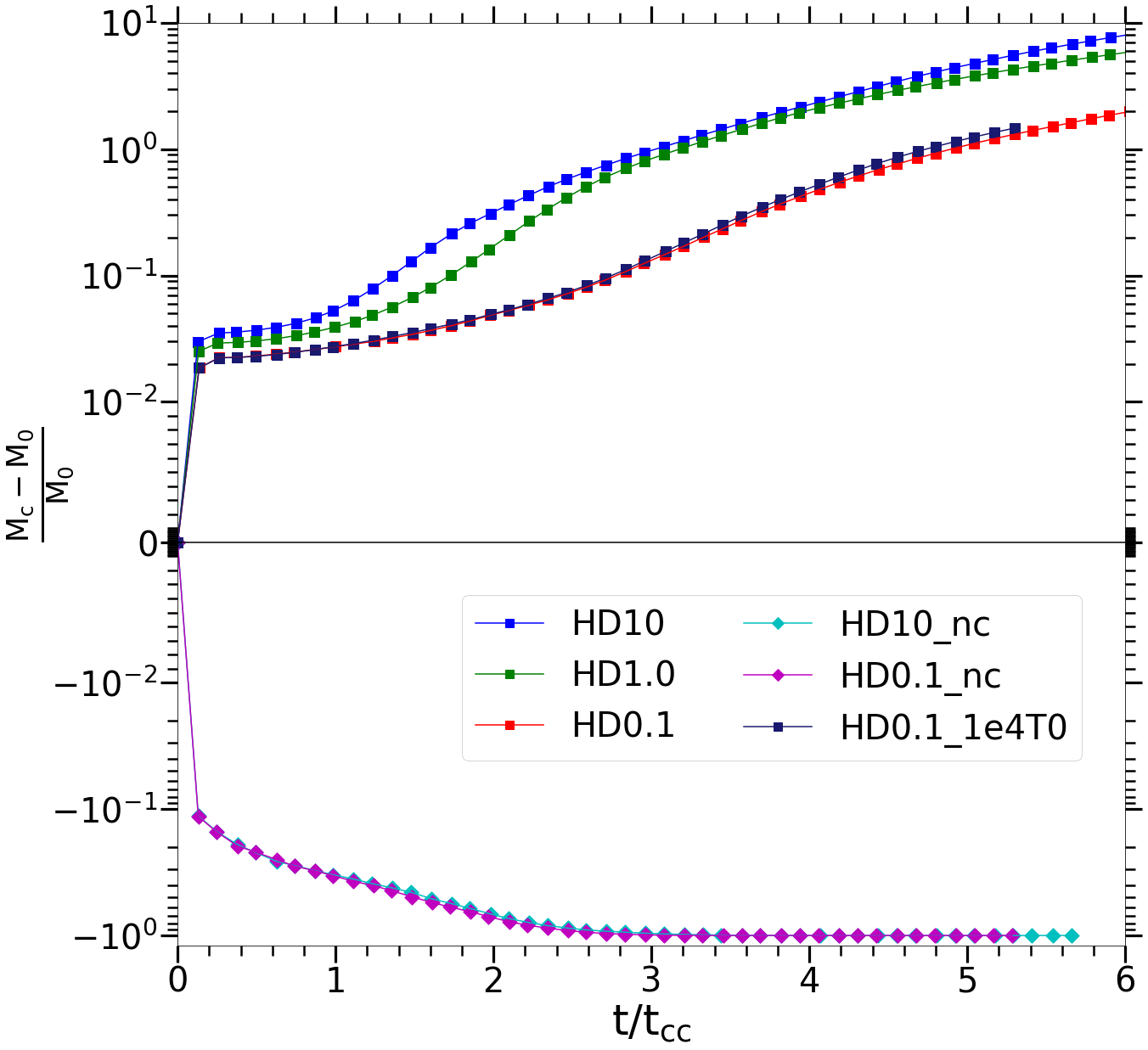}
    \caption{The fractional change in cold gas mass shown for cooling and non-cooling simulations at different cloud densities. The cloud is destroyed within a few \tcc\, if cooling is not present consistent with a large body of previous literature. However, in radiative simulations significant cloud mass growth is seen for all densities explored.}
    \label{plot:coolvsnocool_cold_mass}
\end{figure}

\begin{figure}
    \centering
	\includegraphics[width=\columnwidth ]{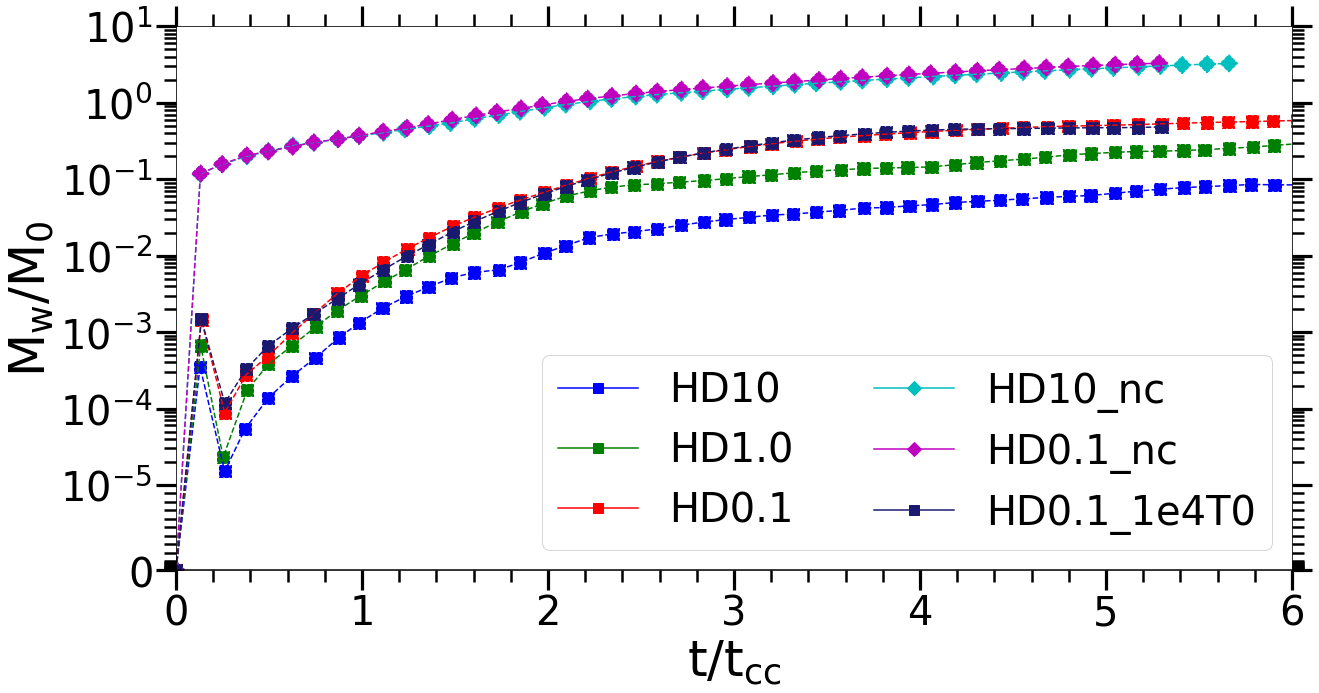}
    \caption{Time evolution of the warm gas fraction normalised to the original cloud mass for different initial densities. In the presence of radiative cooling, the amount of warm gas is reduced in comparison to the non-cooling runs as the warm gas is depleted by gas cooling into the cold, dense phase. However, warm gas gradually builds up over time, with initially more diffuse clouds producing a larger fraction of warm gas. }
    \label{plot:coolvsnocool_warm_mass}
\end{figure}

\begin{figure*}
	\includegraphics[width=18cm]{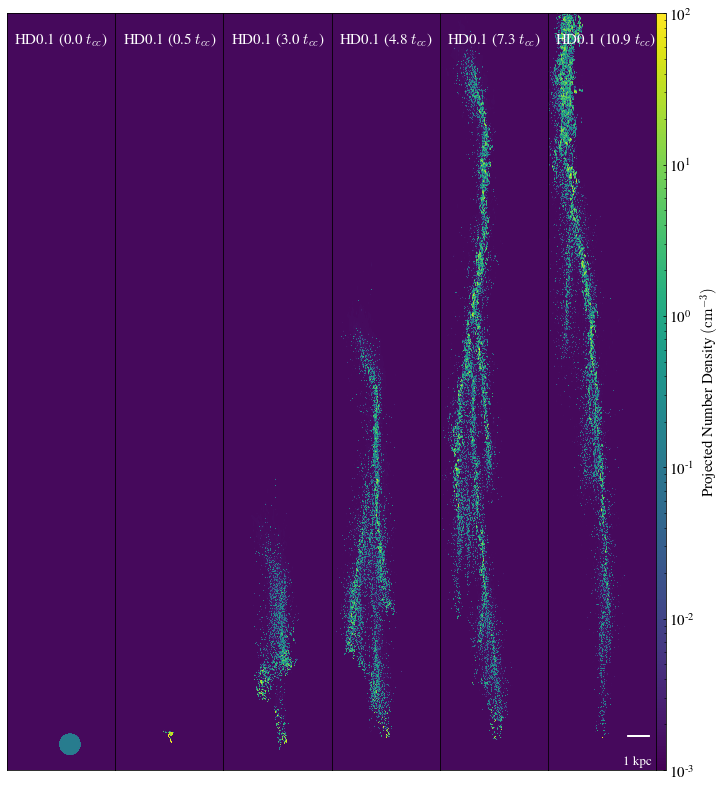}
    \caption{Density-weighted projections of the number density $n$ shown at different times for simulation \run{HD0.1}. The cloud collapses and fragments within one crushing timescale, and the resulting clumplets then interact dynamically with the wind, forming an extended tail. Note that each projection is centred on the most upstream clump material.}
    \label{plot:hydro multi-slice}
\end{figure*}

\begin{figure}
    \centering
    \includegraphics[width=\columnwidth]{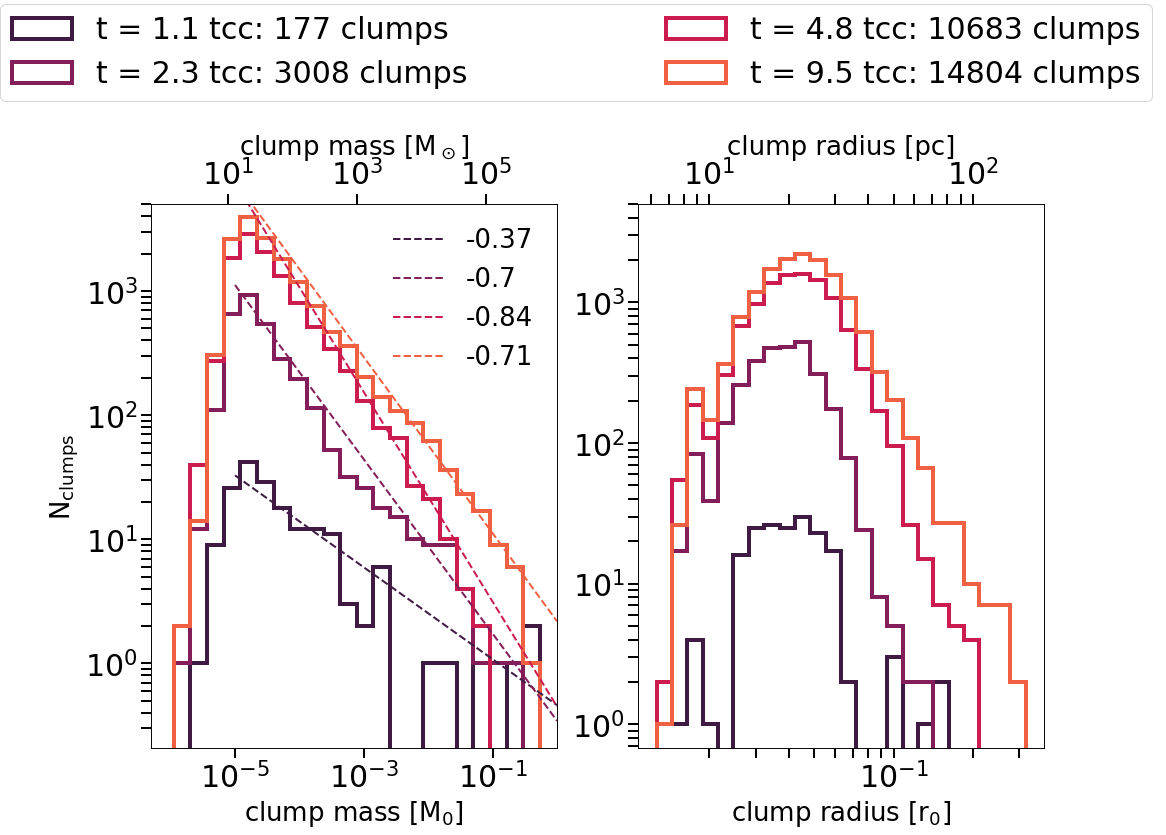}
    \caption{Distribution of clump masses (left) and clump radii assuming spherical clumps (right) for simulation \run{HD0.1} at four different epochs. Insets in the left panel show the slopes of the power law fit for clump masses above $10^{-5} \rm \ M_\odot$. }
    \label{plot:clump_stats_HD01}
\end{figure}

The cloud crushing timescale for the clouds we use is $24.4$ Myr, and the cooling timescale is initially a factor $10^{4.6}$ shorter than this for the densest cloud \run{HD\_10}, and $10^{3.6}$ and $10^{2.6}$ times shorter than the crushing time for \run{HD\_1.0} and \run{HD\_0.1}, respectively.
As shown in Fig.~\ref{plot:HD0.1_nc vs HD0.1 density}, cooling completely changes the morphology of cold clouds. Without cooling (left panel), cloud gas is mixed rapidly into the hot phase, while with cooling (right panel) the cold cloud is shattered into many small fragments that persist. More quantitatively, Fig.~\ref{plot:coolvsnocool_cold_mass} shows that the clump in the \run{HD0.1\_nc} run loses 70\% of its initial mass by $t=2 \,t_{\rm cc}$, which rises to 99\% by $t=3 \,t_{\rm cc}$. The cloud being destroyed on a timescale of the order of \tcc\ is in agreement with standard results of the cloud-crushing problem. This evolution is independent of the initial clump density that we vary by two orders of magnitude (compare the low density non-cooling cloud \run{HD0.1\_nc} to the high density non-cooling cloud \run{HD10\_nc} in Fig.~\ref{plot:coolvsnocool_cold_mass}).

When radiative cooling is included, the cloud follows a completely different evolution, as shown in Figs.~\ref{plot:coolvsnocool_cold_mass} and \ref{plot:coolvsnocool_warm_mass}, which quantify the time evolution of the cloud and wind mass, respectively, and visualized in Fig.~\ref{plot:hydro multi-slice} which illustrates the time evolution of the projected number density. Most notably, rather than being destroyed, the total mass of the cloud grows significantly over time. This long-term cloud growth is expected from previous work by \citet{Li2020} and \citet{Gronke_Peng_2020}, who both predict a minimum cloud survival radius (of $6.5$ and $3.3$~pc, respectively, computed using Equations~(3) and (5) from \citealp{Kanjilal2021}) well below the initial $500$~pc radius of the cloud studied here. 

Initially, for $t \lesssim $ \tcc \ the cold mass in the clouds grows slowly, but then it accelerates, doubling within 5 \tcc\ even for the lowest initial cloud density explored (see Fig.~\ref{plot:coolvsnocool_cold_mass}). This cold mass growth is more pronounced for the runs with higher initial cloud densities (see Section~\ref{sec:Density Dependence} for a more detailed discussion) and for simulations performed at higher resolution (see Appendix~\ref{appendix: convergence}).

In Fig.~\ref{plot:clump_stats_HD01}, we investigate whether this increased growth is due to a larger number of individual clumps, or an increased mass per clump. To do so, we identify individual cloudlets in the wake as described in Section~\ref{sec:clump_finding} and then study the statistics of this ensemble of cloudlets for each simulation over time. As can be seen in Fig.~\ref{plot:clump_stats_HD01}, the total growth in cloud mass is both due to a higher number of individual clumps, and due to a higher mass per clump.

The process commences with an initial collapse phase, during which the cloud cools rapidly and compacts on sub-crushing timescales. Density inhomogeneities are seeded in the collapse, and material starts to ablate from the main cloud in the form of individual fragments with radii in the range $5-100$ pc, as shown in Fig.~\ref{plot:clump_stats_HD01}. Over time, the original cloud is entirely broken up and the resulting clumps, as well as the warm, diffuse material between them, forms an extended clumpy wake, as shown in Fig.~\ref{plot:HD0.1_nc vs HD0.1 density} and the late-time panels of Fig.~\ref{plot:hydro multi-slice}. The typical slope of $dN_{\rm clumps}/dM$ is of the order of -0.7 (see fitted slopes in Fig. \ref{plot:clump_stats_HD01}), which is somewhat shallower than the -1 measured in \citet{Gronke2022}, which suggests that the higher levels of turbulence in \citet{Gronke2022} accelerate shattering. We find, by analysing the typical range of clump sizes and densities seen post-shattering, that the characteristic crushing timescale of the clumps is roughly the same as that of the initial parent cloud, allowing us to continue use of \tcc \ as a robust characteristic time even after the shattering process. The overdensity $\chi$ does however increase, to approximately $100 \times T_{cl,0} / T_{cl,ps}$ for our initial overdensity of $100$ and a post-shattering temperature $T_{cl,ps}$ when the cloud attains approximate hydrostatic equilibrium again.\\
We run one cloud in initial hydrostatic non-equilibrium, \run{HD0.1\_1e4T0}, by setting the initial cloud temperature to $T_{cl} = 10^4$~K, to examine how sensitive the shattering and further evolution is to the initial temperature and pressure of the clump. As can be seen in the mass evolution plots of Fig \ref{plot:coolvsnocool_cold_mass} and \ref{plot:coolvsnocool_warm_mass}, shattering occurs quickly enough that evolution is insensitive to the initial cloud temperature, and the only cloud property we find to be important is the density.

 \begin{figure*}
	\includegraphics[width=\textwidth ]{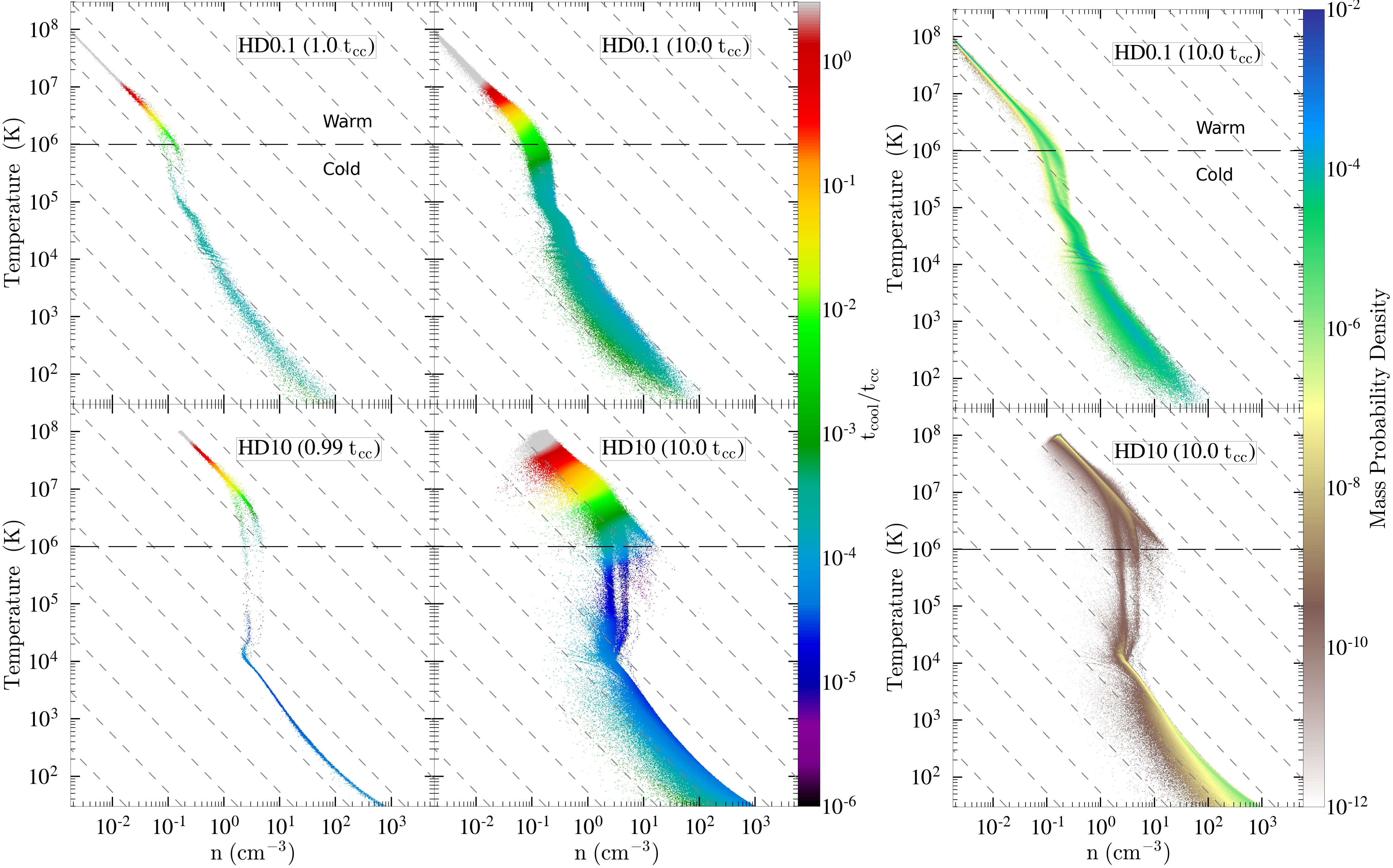}
    \caption{\textbf{\textit{Left:}} Binned temperature-density phase space occupied by the clouds for simulations \run{HD0.1}, and \run{HD10}, with the colourmap showing the ratio of cooling time to crushing time $\theta$ at three different times (left to right). We also show the regions we define throughout the paper as 'warm' gas ($10^6<T<10^8$ K) and 'cold' gas ($T<10^6$ K). The change in initial cloud density (as well as wind density) leads to a different phase space region of the cooling function being sampled by the cloud, particularly in the warm phase; at higher densities the cooling times are much shorter, and cooling can take place as much as nearly six orders of magnitude faster than cloud crushing for the densest cloud we test. The densest cloud \run{HD10} cools isochorically in the warm phase by approximately one and a half orders of magnitude more than the least dense cloud, and settles later onto the isotherm that will take the temperature down to $T_{\rm fl}$. \textbf{\textit{Right:}} We instead show the mass probability density of the two clouds each after 10 \tcc. Though the initial overdensities were identical, the more dense cloud \run{HD\_10} holds more of its mass in dense, cold fluid, and has a more bimodal distribution between the cold and hot phase, whereas \run{HD\_0.1} has a mass distribution more evenly distributed through the warm phase as well.}
    \label{plot:phaseplot}
\end{figure*}

The late-time evolution after several crushing times is driven by the evolution of the wake in the cooling case. The mixing of the cold fragments with the hot \ac{icm} generates a warm, mixed phase, which makes up the wake of the clump, and trails any high-density remnants of the original clump that remain at the 'head' of the cloud. The amount of gas contained in this wake grows over time but remains small in comparison to the cold gas mass, reaching about about $25\%$ of the original cloud mass by $t=3 t_{\rm cc}$ for \run{HD0.1} (see Fig.~\ref{plot:coolvsnocool_warm_mass}). This warm gas has a very short cooling time, as can be seen in Fig.~\ref{plot:phaseplot}, due to a peak in the cooling function $\Lambda(T)$ between $10^5\, \rm K$ and $10^6\,\rm K$. In the most extreme cases, the cooling time can be $5$-$6$ orders of magnitude shorter than the crushing timescale. For such a short cooling time, cooling dominates the evolution of the gas, and gas transfers quickly through the temperature-density phase space from high temperature, diffuse to cold, newly condensed cloud\footnote{We note that the bimodal feature in the temperature range $10^4 - 10^6 \rm \ K$ in Fig.~\ref{plot:phaseplot} is due to the structure of the flow, and results from a combination of cooling and mixing at the surface of individual cold clumps. The left-hand feature is formed by gas up-stream of individual clumps, and consists predominantly of gas that originates in the hot wind. The right-hand feature is composed of gas down-stream of individual clumps, which contains a higher percentage of gas ablated from the clumps and is therefore at higher density.}. The mixed phase condenses onto the clump as \textit{new} cold gas, which means that the cold fragments act as nucleation sites for cold condensate within the hot \ac{icm}. We find that this is dominant over mixing losses, resulting in the growth of the total cold gas mass over time (see Fig.~\ref{plot:coolvsnocool_cold_mass}). The no-cooling cases, display no such growth, even though a significant amount of warm gas is produced, with the fraction of warm gas more than an order of magnitude larger in the no-cooling runs in comparison to the initial cloud mass (see Fig.~\ref{plot:coolvsnocool_warm_mass}). This markedly different evolution with cooling shows the importance of including radiative cooling in driving the evolution of cold clouds in hot atmospheres of galaxy clusters.
 
Qualitatively,  our results on cold cloud growth agree with recent results by \citet{Jennings_michael_2020}, \citet{Gronke_Peng_2020} and \citet{Li2020}, as well as with results by \citet{Gronke_2018} even though the latter do not see the same amount of shattering as they impose a higher temperature floor of $4\times 10^4 \rm \ K$ (see Section~\ref{sec:Cooling Floor} for a discussion on the impact of cooling floors). The cold clumps simulated here exceed the pressure condition required for shattering put forth by \citet{Gronke_Peng_2020}, namely $P_\text{cl,ps}/P_\text{cl,0} \sim T_\text{cl,ps}/T_\text{cl,0} \lesssim 0.33$, where $T_\text{cl,ps}$ is the post-shattering cloud temperature, equal to the cooling floor temperature. We do not find any significant post-shattering coagulation of fragment clumps, probably due to the extreme pressures generated by the large temperature gradients, with the fragments being well-separated after the post-shattering expansion. This can be clearly seen at late times in Fig.~\ref{plot:hydro multi-slice}.
 
The final size of clumplets needs to be considered in this context too. \citet{McCourt_2017} put forth an argument that clouds will shatter until their clumps are small enough that the sound-crossing time is of order the cooling time, and thus cooling can occur isobarically. This occurs when clumps reach a physical lengthscale of $l \sim c_{\rm s}t_{\rm cool}$, which occurs approximately on the timescale of $t_{\rm cool}$. \citet{McCourt_2017} show that clouds should shatter down to a size of $l_\text{cloudlet} \sim c_{\rm s}t_{\rm cool} \sim (0.1 \text{ pc})/n$, and this has been further confirmed by other simulation works \citep[see e.g.][]{Gronke_Peng_2020}. The largest characteristic length of these cloudlets in our simulations is expected to be around a parsec, an order of magnitude smaller than our minimum resolved scale, even with the high resolution we have adopted. Ideally, a resolution two order of magnitude better than our simulation runs should be used, to investigate whether shattering stops at this characteristic scale, or if clumps will fragment even more, but this is computationally unfeasible with our current set-up since $r_{\rm 
 cl}/l_\text{cloudlet} \gg 1$. 
 
In summary, we reproduce the destruction or growth of the cloud on crushing timescales found by similar studies. For the parameter space of cold clouds in a hot cluster environment, we find that clouds grow in mass on timescales relevant for their trajectories when radiative cooling is taken into account. This growth is significant, and leads to cold clouds acting as nucleation sites within the hot \ac{icm}, with condensation beginning after only a few crushing timescales, which in cluster environments translates to a few tens of Myr. 

\subsubsection{Density Dependence} \label{sec:Density Dependence}

We now compare the effects of altering the initial cloud density on the evolution of clouds themselves, by comparing runs with initial cloud densities of $\rho_{\rm cl} = 10 \ {\rm cm}^{-3}$ (\run{HD10}) and $\rho_{\rm cl}=1 \ {\rm cm}^{-3}$ (\run{HD1.0}) to the fiducial run \run{HD0.1} which has cloud density $\rho_{\rm cl}=0.1 \ {\rm cm}^{-3}$. All three simulations have the same  overdensity $\chi = \rho_{\rm cl} / \rho_{\rm w} = 100$, and are therefore identical within the context of scale-free considerations. 

As expected from analytic work, the absolute density of the cloud does not change the mass evolution in the absence of cooling as long as $\chi$ is constant. This can be clearly seen by the identical time evolution of both the warm and cold gas for simulations \run{HD10\_nc} and \run{HD0.1\_nc} in  Fig.~\ref{plot:coolvsnocool_cold_mass} and Fig.~\ref{plot:coolvsnocool_warm_mass}.

It is only with the addition of cooling that the absolute density of the cloud makes a difference to its evolution. An initially lower density cloud forms lower density fragments post-shattering as would be expected, and the lower density of the clumplets in the wind results in greater mixing of cold gas into a warm phase. This can be clearly seen in Fig.~\ref{plot:coolvsnocool_warm_mass} by the larger amount of warm gas generated by the lower density cloud, in comparison to its initial mass. This confirms that the presence of cooling breaks the scale-free nature of the problem, by introducing an additional characteristic timescale which depends on both the cloud density and temperature (as well as metallicity, but this is kept constant in our runs). As the cooling time is shorter for denser material, initially denser clouds show faster mass growth due to the more rapid condensation of gas from the warm to the cold phase. The impact of the higher density mixed gas on the cooling time for all three simulations can be seen in Fig.~\ref{plot:phaseplot}, with the fastest cooling regime in the warm phase only occupied by the densest clouds. Interestingly, all three simulations reach a rough equilibrium between mixing new gas into the warm phase, and cold gas condensing into the cold phase, but the equilibrium amount of warm gas is lower for initially higher density clouds (see Fig.~\ref{plot:coolvsnocool_warm_mass}). 

For our densest cloud modelled here, \run{HD10\_nc},  the initial cloud conditions are such that \citet{Li2020} predict that clouds with an initial radius of $r_{\rm cl,init} < 651\rm \ pc$ \citep[computed using Equation~(5) in][]{Kanjilal2021} should not survive, which is larger than our initial cloud radius of $500$~pc. By contrast, \citet{Gronke_Peng_2020} predict a minimum cloud survival radius of $334$~pc \citep[computed using Equation~(3) in][]{Kanjilal2021}, and therefore predict cloud \run{HD10\_nc} should survive. With only a single cloud in this borderline region between the two regimes, we refrain from strong statements about either criterion. Instead, we simply note that \run{HD10\_nc} exhibits robust growth in our simulations and that this range of radii would be an interesting range to probe in future work.

We conclude that cooling is the dominant mechanism determining the evolution of the cloud for all clouds within the two orders of magnitude in density probed here. We also confirm that in cluster environments, where cooling dominates the evolution of cold clouds, scale-free studies are insufficient and the parameter space needs to be explored widely. The smaller the ratio of cooling time to cloud crushing time $\theta$ (see Equation~(\ref{eqn: ratio of timescales})), the faster cold clumps in the \ac{icm} can condense new gas and grow, with all cases tested here more than doubling the clouds initial mass on timescales of around $150 \rm \ Myr$, and some increasing it by almost a factor $10$.

\subsubsection{Cloud crushing and small scale heating} \label{sec:Cooling Floor}

The sub-kpc scale physics which takes place in cold clouds within galaxy clusters is poorly understood. However, observations show the presence of warm gas, despite short cooling times, which suggests that clouds need to be persistently powered. Over the years, many processes have been suggested as the source of this heating, from photoionization by the central \ac{agn} \citep[e.g.][]{Heckman1989, McNamara2012} or massive stars \citep{Canning2014}, via thermal conduction \citep{Voit2008} and magnetic reconnection \citep{Hanasz1998,Tanuma2003, Churazov2013} to cosmic ray streaming heating \citep{Ruszkowski2018}, to name a few. Stars can form in ``knots'' along filaments \citep{Vantyghem2018}, and these sites may also provide some heating and feedback to the cold gas once fragmented. So far, no consensus has been reached on the origin of cloud heating, but such energy injections at the cloud radius would off-set some of the cooling and prevent clouds cooling below an effective temperature floor. This floor would of course be expected to be time-varying, with the characteristic timescale of the most dominant heating mechanism setting the timescale of variation. However, without knowing where the dominant energy input comes from, we choose to keep the floor constant throughout our simulations.

In the simulations presented here, we test the impact of such small-scale heating and feedback processes, as well as the heating effects of a radiation background within the cluster, by varying the cooling floor, i.e. the minimum temperature that gas can reach through radiative cooling. The implicit assumption is that at this temperature, the energy losses from radiative cooling match the combined heating effects of the processes we do not explicitly model. To test the robustness of our general results against this choice of cooling floor, we select one simulation (\run{HD0.1}) to be repeated with a variety of cooling floor temperatures \Tfloor. These runs are detailed in Table~\ref{tab:hydro runs}.

\begin{figure}
	\includegraphics[width=\columnwidth ]{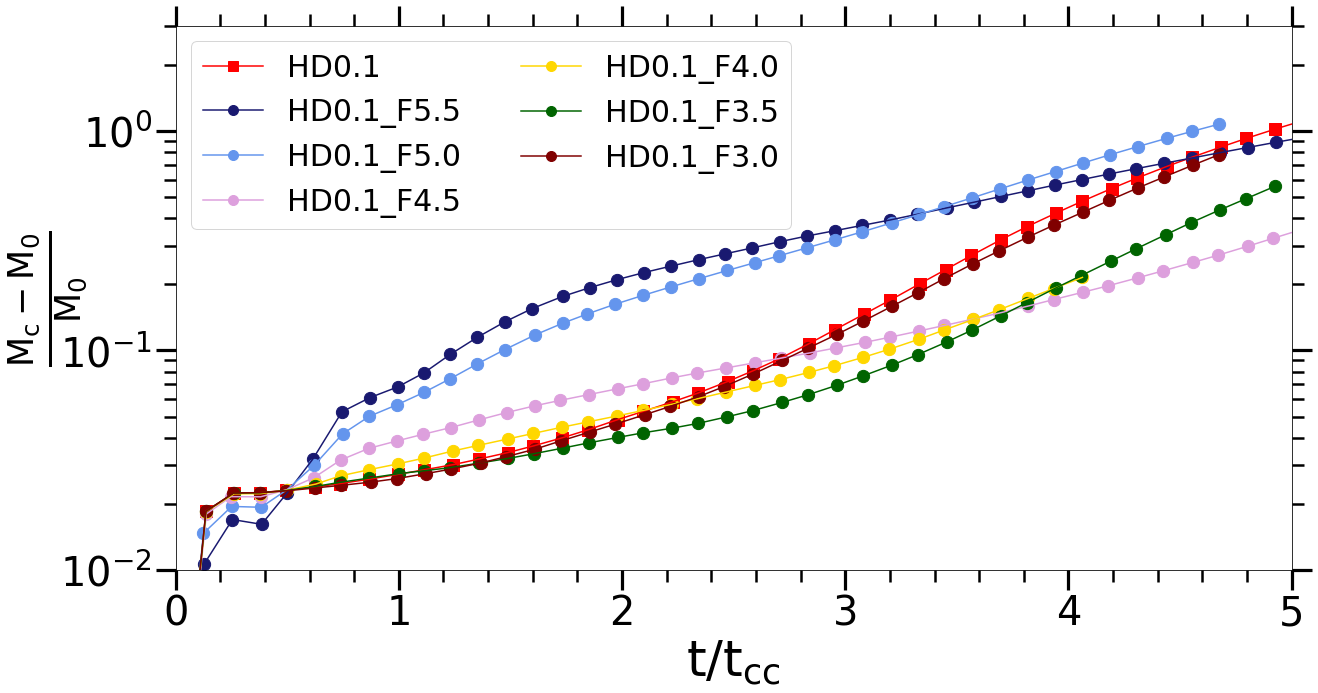}
    \caption{The normalised cold gas mass fraction for variations of \run{HD0.1} run with different cooling floor temperatures $T_{\rm fl}$. They are compared to our flagship run \run{HD0.1}, which has an effective temperature floor of $T_{\rm fl} = 10\rm \ K$. Trends are non-linear, with cold gas mass growth the fastest for both high ($T_{\rm fl} \geq 10^5 \rm \ K$) and low ($T_{\rm fl} \leq 10^3 \rm  K$) cooling floors, but suppressed for $10^3 < T_{\rm fl} < 10^ 5 \rm \ K$.}
    \label{plot:cooling_floor_cold}
\end{figure}
\begin{figure}
    \centering
	\includegraphics[width=\columnwidth ]{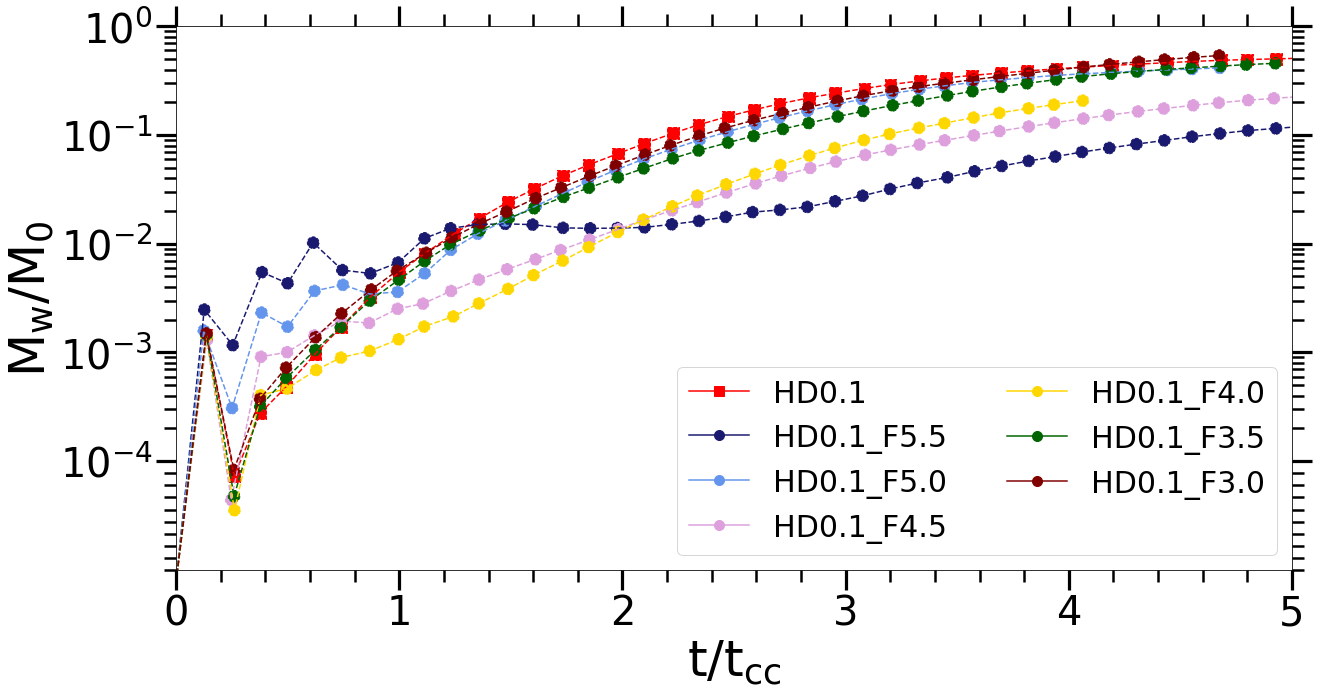}
    \caption{The normalised warm gas mass fraction for each variant of \run{HD0.1} run with different cooling floor temperatures $T_{\rm fl}$. Lower cooling floors produce more warm mass in the temperature range $10^3 \ {\rm K} \leq T_{\rm fl} < 10^{5.5} \ {\rm K}$, but significantly lower cooling floors again reduce the warm gas mass.}
    \label{plot:cooling_floor_warm}
\end{figure}

We find that clouds grow for all choices of cooling floor tested here. During the initial cloud disruption phase ($t < 0.5 \ t_{\rm cc}$), a higher cooling floor produces less cold gas (Fig.~\ref{plot:cooling_floor_cold}) and more warm gas (Fig.~\ref{plot:cooling_floor_warm}). The cloud with a cooling floor at $3 \times 10^5$ K hardly shrinks in radius during $t<$ \tcc, as the initial cooling and resulting loss of pressure is strongly reduced, while clouds with lower cooling floors contract significantly as can be seen in the images in Fig.~\ref{plot:hydro multi-slice} for our fiducial \run{HD0.1} run. 

Beyond 1 \tcc, the evolution becomes more complex. There is a general trend for higher cooling floors to produce more cold gas for $0.5$ \tcc $< t < 4 $ \tcc, while for lower cooling floors the cold gas mass increase is lower initially but then accelerates after several crushing times. The lower the cooling floor, the faster the turn-up occurs, but results converge for minimum temperatures of $T_{\rm fl} \leq 10^3 \rm \ K$. As a result of this speed-up in condensation for low values of $T_{\rm fl}$, by 5 \tcc, the highest and lowest $T_{\rm fl}$ have a similar amount of total cold gas mass, while the intermediate-floor runs have a smaller amount.

The evolution in warm gas mass, shown in Fig.~\ref{plot:cooling_floor_warm} is similarly non-linear. Again the early evolution is directly dependent on the cooling floor temperature $T_{\rm fl}$, with lower $T_{\rm fl}$ producing less warm gas for $t<$ \tcc. For later times, the long-term evolution of the total warm gas mass $M_{\rm w}$ is generally a function of $T_{\rm fl}$, with higher $T_{\rm fl}$ producing less $M_{\rm w}$. The outlier to this trend is \run{HD0.1\_F5.0}, which despite having the second highest $T_{\rm fl}$ produces a similar amount of $M_{\rm w}$ as runs with $T_{\rm fl} \leq 10^{3.5} \rm \ K$.

\begin{figure}
    \centering
    \includegraphics[width=\columnwidth]{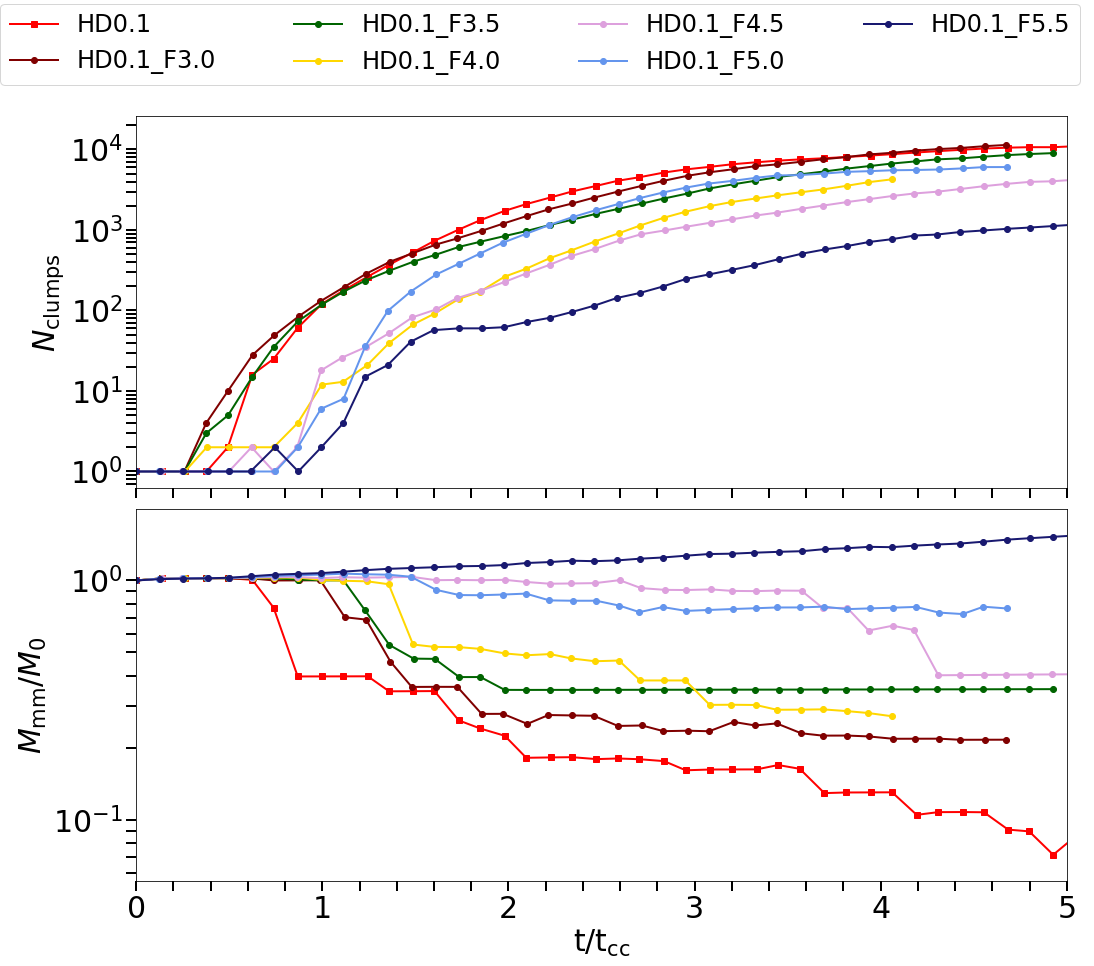}
    \caption{Time evolution of the number of clumps (top panel) and the mass of the most massive clump as a function of initial cloud mass (bottom panel) for variations of the \rm{HD0.1} run with different $T_{\rm fl}$. In general, higher minimum temperatures lead to less fragmentation and a more stable main clump but the evolution is not linear with $T_{\rm fl}$.}
    \label{plot:cooling_floor_clumps}
\end{figure}

This somewhat complex behaviour in the warm phase is due to how cloud shattering proceeds for each simulation. As can be seen in Fig.~\ref{plot:cooling_floor_clumps}, the cloud evolution changes fundamentally with increasing $T_{\rm fl}$. For low values of $T_{\rm fl}$ we see an evolution very similar to that shown for \run{HD0.1} in Fig.~\ref{plot:hydro multi-slice}: the original cloud shrinks rapidly very early on as it loses pressure support, and breaks up into a large number of individual clumps. During this time, the total cold gas mass increases, but the mass of the most massive clump $\rm M_{\rm mm}$ actually reduces over time as shattering progresses. This occurs within half a crushing timescale and produces small clumplet fragments, which interact with the wind over the next few crushing times to produce individual, but often connected, wakes of warm gas. This process has been found recently in other studies too, notably by \citet{McCourt_2017}, and \citet{Gronke_Peng_2020}, though \citet{Jennings_michael_2020} notes that it is still uncertain as to whether fragmentation is shock-induced due to the Richtmyer-Meshkov instability \citep{meshkov1972}, or if some other process is at play, such as pressure interactions between two neighbouring clouds.

For higher $T_{\rm fl}$, the shattering is less strong and the number of individual clumps remains much lower. This produces less mixed gas before $0.5$ \tcc \ as a result of hot gas not being pulled violently into a mixing layer by the pressure gradients during the collapse. In general, higher $T_{\rm fl}$ means a more intact main clump with a higher mass (see evolution of the mass of the most massive clump, $M_{\rm mm}$, in Fig.~\ref{plot:cooling_floor_clumps}). For the highest values of $T_{\rm fl}$ tested here, the main clump remains intact and actually grows in mass over time. The cooling floor temperature has such an important impact because the initial cooling of the cloud, once placed in the \ac{icm}, occurs rapidly enough that it is essentially isochoric, which means that the pressure gradient produced at the boundary is linearly proportional to the value of the temperature floor where the cloud settles, and so there is a range of orders of magnitude between the forces these clumps experience across runs. We also note that the number of clumps after the initial ablation event is closely linked to $\rm l_{\rm cloudlet}$ shown in Fig.~\ref{plot:shattering lengthscale}: runs with $T_{\rm fl}$ corresponding to maxima of $\rm l_{\rm cloudlet}$ (for example \run{HD0.1\_F5.5} or \run{HD0.1\_F4.0}) have less clumps early on than those corresponding to local minima (such as \run{HD0.1\_F4.5} or \run{HD0.1\_F5.0}).

After reaching an initial equilibrium, the clouds at the higher temperature floor tend initially to \textit{ablate} more than mix, with clumplets breaking or shredding off from the main clump, increasing the effective mixing surface, and generating mixed gas at the boundary layer. This is perhaps a more traditional cloud-crushing scenario, and one that approaches the behaviour in the no-cooling limit case. Mixed gas does condense to form new cold gas, but the physical extent of the wake at very high temperature floors appears to be suppressed, and there is a larger consolidated ``head'' of the cloud that remains in the wind. 

The amount of ablation and shattering is directly related to the amount of warm gas produced, as can be seen by studying the outlier simulation \run{HD0.1\_F5.0}, which produces many more individual clumps around 1.5 \tcc \ than either \run{HD0.1\_F5.5} or \run{HD0.1\_F4.5}. This early ablation event is also the exact point in time when the warm gas mass $\rm M_{\rm w}$ for this simulation increased significantly in Fig.~\ref{plot:cooling_floor_warm}.

We conclude that internal heating mechanisms could play an important role in the long-term evolution of cold clouds by suppressing ablation and shattering, and allowing the main cloud to survive for longer. However, in a cluster environment, mixing at the cloud surface remains sufficiently efficient that even with reduced shattering clouds grow in mass over time, albeit slower than those that originally were originally mixed more efficiently into the warm phase. 

\subsubsection{Kelvin-Helmholtz vs Pressure Gradients} \label{sec:KH vs Pressure Gradients}

The generation of a mixed phase at the boundary layer of the cloud is the key process that drives the evolution and long-term growth of the clouds via condensation in our simulations. \citet{Gronke2019}, who find robust cloud growth in a galactic setting, argue that rapid cooling can generate pressure gradients large enough that mixing in the boundary layer is dominated by pressure-driven mixing rather than by Kelvin-Helmholtz processes in the shear layer. If pressure mixing dominates, the cooling rate directly determines the growth rate of the cloud, and growth continues even when the cloud is almost co-moving. 

\begin{figure}
	\includegraphics[width=\columnwidth ]{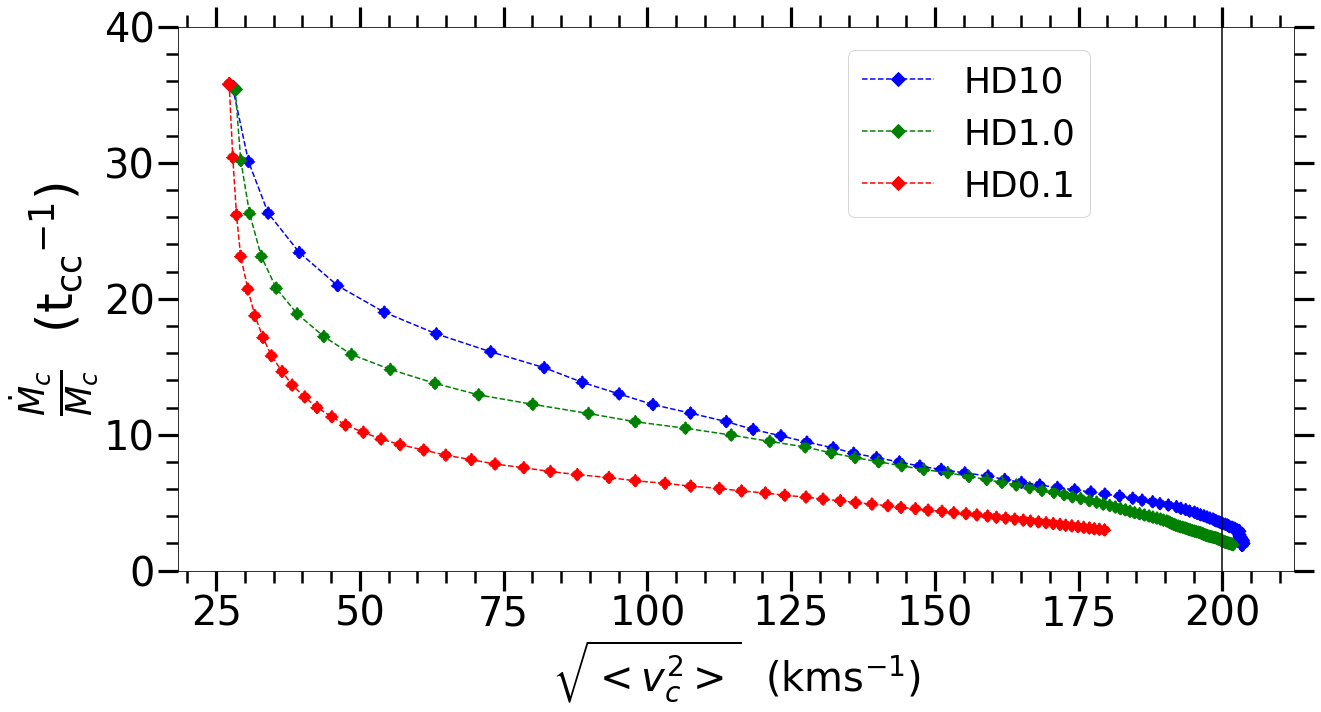}
    \caption{The root mean square velocity of the cold gas against the growth rate of the cold cloud, normalised to the total cold mass in the box at the present time for each data point, in units of $t_{cc}^{-1}$, for our three hydro runs. The initial collapse and shattering phase is not included. Note that the mean square velocity of the cloud relative to its \textit{initial} zero velocity is plotted, and that the wind velocity sits at $200$ km\,s$^{-1}$. The mass-normalised growth rate reduces as the cloud becomes co-moving in the wind, indicating that the mixing is largely Kelvin-Helmholtz instability-driven and that the rate of condensation onto the cloud reduces as the cloud decelerates and the Kelvin-Helmholtz mixing becomes less effective.}
    \label{plot:pressure_vs_KH_mass_rate}
\end{figure}

To understand how mixing proceeds in our simulations, we plot the growth rate of the cold gas fraction normalised to the current cold mass ${\dot{\mathrm{M}}\mathrm{_c/M_c}}$ against the velocity of the cloud (which is initially at rest in the simulation frame) in Fig.~\ref{plot:pressure_vs_KH_mass_rate}. Note that `cloud' refers to the entire wake of mixed gas, not the most massive remnant of the original cloud. As can be seen in Fig.~\ref{plot:cooling_floor_clumps}, the most massive clump looses mass over time, and therefore has a \emph{negative} growth rate, while the cloud has \emph{positive} growth rate as the total gas mass increases over time. Random motions not along the axis of flow are included in this root-mean-square velocity, since this velocity dispersion can still contribute to instability mixing, although in practice the velocity magnitudes are in general small compared to the bulk flow. (This can be seen by only a small excess velocity of the gas in the approximately co-moving state reached by \run{HD1.0} and \run{HD10}.)

We find that though the total cloud growth rate continues to increase with time as the cloud becomes approximately co-moving with the wind, the normalised rate -- which takes into account the fact that at later times the cloud is larger and has a greater mixing area -- reduces. For example, when the overall cloud reaches approximately co-moving status, the normalised growth rate is around $5-10\%$ the value at a relative velocity of 170 km\,s$^{-1}$.
This supports the argument that the condensation and cloud growth is driven by Kelvin-Helmholtz mixing, which acts to a lesser extent when the cloud slows down relative to the wind, rather than the pressure mixing. In the growth rate of the warm fraction (not shown), we see a similar decline as the cloud becomes co-moving, but a peak in growth rate between $75$ and $100$ km\,s$^{-1}$; most likely this corresponds to the point where the cold gas area becomes completely covered in a warm mixing layer. 

In clusters this slowing down occurs due to both the ram pressure experienced by the cloud gas as it punches through the \ac{icm}, and also due to the gravitational effect of the cloud climbing out of the central potential in the cluster. The cloud can also slow down/become entrained due to cooling gas imparting momentum to the cloud as it condenses. Treating this case as an inelastic collision and conserving moementum, the approximate velocity relative to the wind at a given time will be $\frac{\Delta v(t)}{v_{wind}} \approx \frac{m_{cl,0}}{m_{cl}(t)}$ \citep{Gronke_2018}.

\subsection{Magnetic fields and cloud growth}
\label{sec:mhd}

Observational studies of galaxy clusters using Fast Radio Burst (FRB) probes \citep[e.g.][]{Prochaska2019} and rotation measures \citep[e.g.][]{Farnes_2017, Malik2020} have shown that the ICM is magnetised with a field strength of about $ 2 \, \rm{\mu}$G. However, field strengths in the centres of cluster cooling flows may be an order of magnitude higher \citep[see][]{Carilli2002}, and recent observations of the Smith cloud (a high velocity cloud around $3$~kpc below the galactic plane of the Milky Way with a size and temperature comparable to our simulated clouds) by \citet{Hill2013} and \citet{Betti_2019} give lower bounds on the peak line-of-sight field strength of $ 8 \, \mu$G and $5 \, \mu$G, respectively. Magnetic fields have long been assumed to be dynamically unimportant in the cloud crushing problem, with the magnetic pressure often orders of magnitude lower than the thermal pressure. However, with mixing being recognised as the dominant mechanism for cloud evolution, more and more simulations of the CGM and ICM have begun to consider their effect in recent years. Such simulations have shown that magnetic fields can help prevent shredding by stabilising perturbation modes aligned with the field (via draping, see e.g. \citealp{MacLow1994,McCourt2015,SparrePfrommer2020}), and by establishing anisotropic thermal conduction \citep[e.g.][]{McCourt2012, Jennings_michael_2020}, with draping then able to suppress conduction across the boundary layer \citep{dursipfrommer_2008}. 

In this section, we investigate how magnetic fields of different strengths and orientations influence the evolution of cold clouds as they travel through the hot ICM. The additional impact of thermal conduction will be studied in Section~\ref{sec:thermal_conduction}. Magnetic fields of initial strength $B$ were initialised in one of two configurations: either \textit{aligned} (``parallel'') with the direction of flow, or \textit{normal} (``perpendicular'') to the direction of flow. We test three different field strengths, with detailed information on parameter choices for magnetised simulations given in Table~\ref{tab:mhd runs}. All simulations presented in this Section and the next are magnetised variations of the \run{HD10} simulation.

\begin{table} 
	\centering
	\caption{The parameters of MHD runs, with columns showing: (1) the names, (2) the magnetic field strength $B_0$, (3) the initial orientation of the magnetic field relative to the initial bulk flow velocity (either normal to the flow or aligned with the flow), and (4) the type of thermal conduction type (if at all). All runs are variations of \run{HD10} and share its parameters where not explicitly stated otherwise.}
	\label{tab:mhd runs}
	\begin{tabular}{lcccr} 
		\hline
    	Name  & $B_0$ [$\mu$G]  & Orientation & Conduction \\
    	\hline
            \run{MHD\_1N}   & 1.0 & normal & \xmark \\
            \run{MHD\_1A}   & 1.0 & aligned & \xmark \\
            \run{MHD\_3N}   & 3.0 & normal & \xmark \\
            \run{MHD\_3A}   & 3.0 & aligned & \xmark \\
            \run{MHD\_5N}   & 5.0 & normal & \xmark \\
            \run{MHD\_5A}   & 5.0 & aligned & \xmark \\
            \run{MHD\_1N\_i}   & 1.0 & normal & isotropic \\
            \run{MHD\_1A\_i}   & 1.0 & aligned & isotropic \\
            \run{MHD\_3N\_i}   & 3.0 & normal & isotropic \\
            \run{MHD\_3A\_i}   & 3.0 & aligned & isotropic \\
            \run{MHD\_5N\_i}   & 5.0 & normal & isotropic \\
            \run{MHD\_5A\_i}   & 5.0 & aligned & isotropic \\
            \run{MHD\_1N\_a}   & 1.0 & normal & anisotropic \\
            \run{MHD\_1A\_a}   & 1.0 & aligned & anisotropic \\
            \run{MHD\_3N\_a}   & 3.0 & normal & anisotropic \\
            \run{MHD\_3A\_a}   & 3.0 & aligned & anisotropic \\
            \run{MHD\_5N\_a}   & 5.0 & normal & anisotropic \\
            \run{MHD\_5A\_a}   & 5.0 & aligned & anisotropic \\
            \hline
    	\end{tabular}
\end{table}

\begin{figure*}
	\includegraphics[width=16cm]{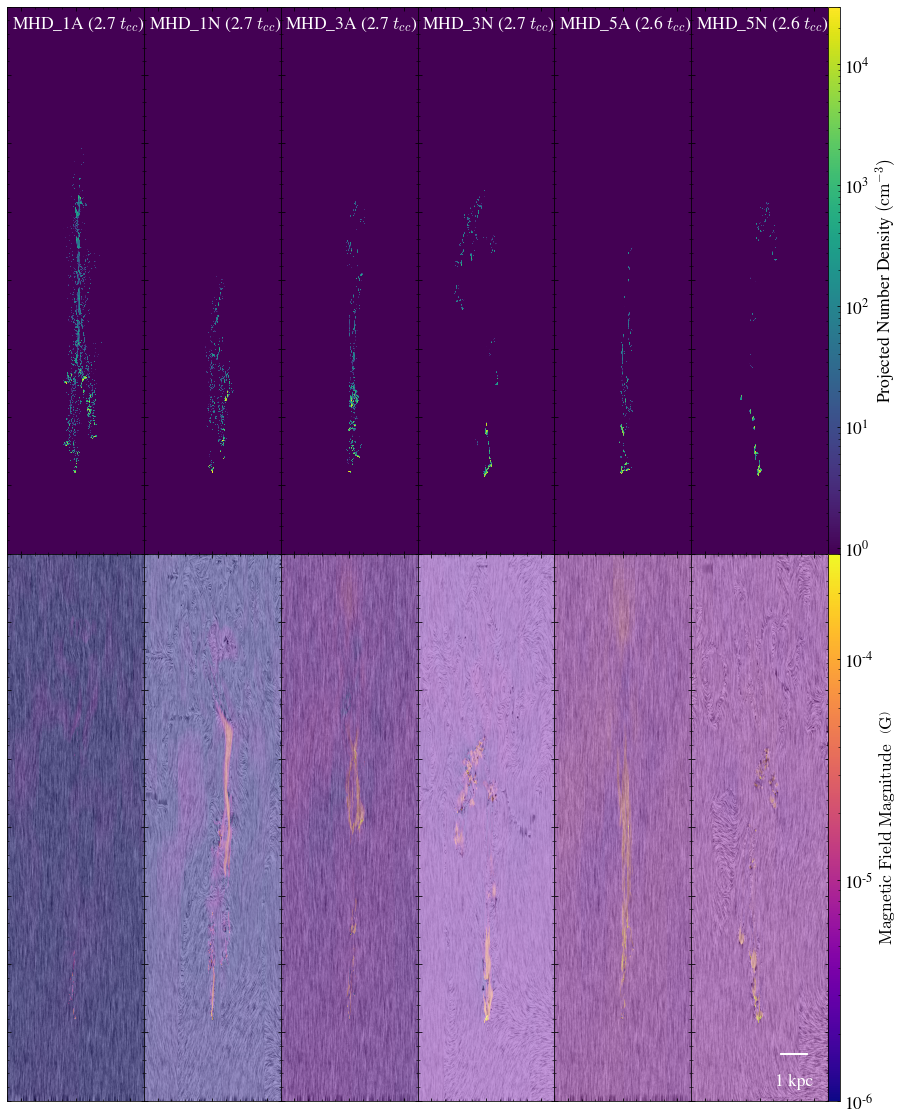}
    \caption{The density-weighted projected number-density (top panel), and the magnetic field magnitude in a slice of the simulation box for the six MHD only runs. On the lower panels the line integral convolution is plotted to give an idea of the orientation of the magnetic field at each region. One can clearly see that if the field is normal to the flow then the field lines drape more around the head of the cloud, and become more non-uniform in the wake. Also, compared to the no MHD simulations at a similar time (see e.g. Fig.~\ref{plot:hydro multi-slice} for comparison, though recall that the initial densities are $100$ times smaller in the pure hydro plot), the amount of dense, cold gas present is highly suppressed, with higher $B$ field leading to more suppression.}
    \label{plot:MHD multislice}
\end{figure*}

\begin{figure}
	\includegraphics[width=\columnwidth ]{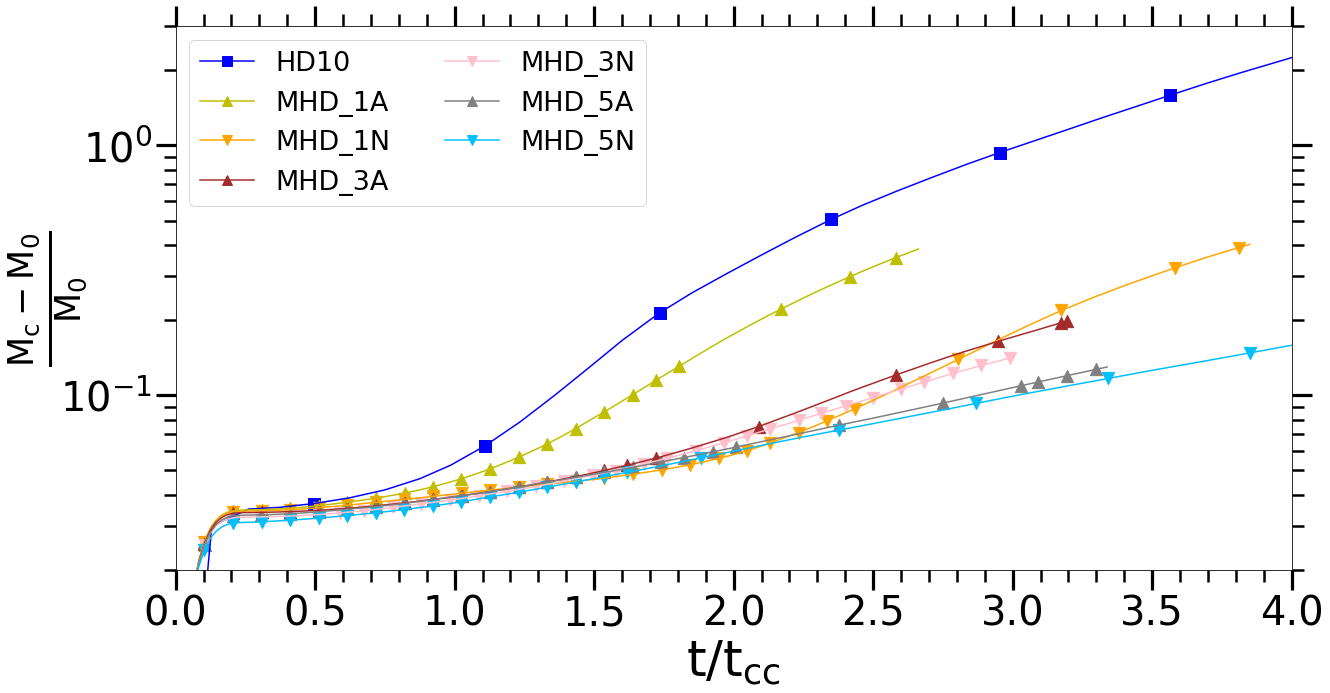}
    \caption{Time evolution of the normalised cold gas mass fraction for six magnetised variations of the \run{HD10} simulation, as well as the corresponding hydrodynamical run. Even a weak field suppresses cloud growth, with suppression being stronger for higher field strengths. Magnetic fields initially perpendicular to the flow seem to be more effective at stunting the cloud growth than initially parallel fields, by a larger factor in the $1 \mu$G case and to a lesser extent for stronger field strengths. Note for that for this plot, and all following magnetic field run mass/curl plots, only every fifth data point is shown for clarity.}
    \label{plot:mhd_cold}
\end{figure}

\begin{figure}
    \centering
	\includegraphics[width=\columnwidth ]{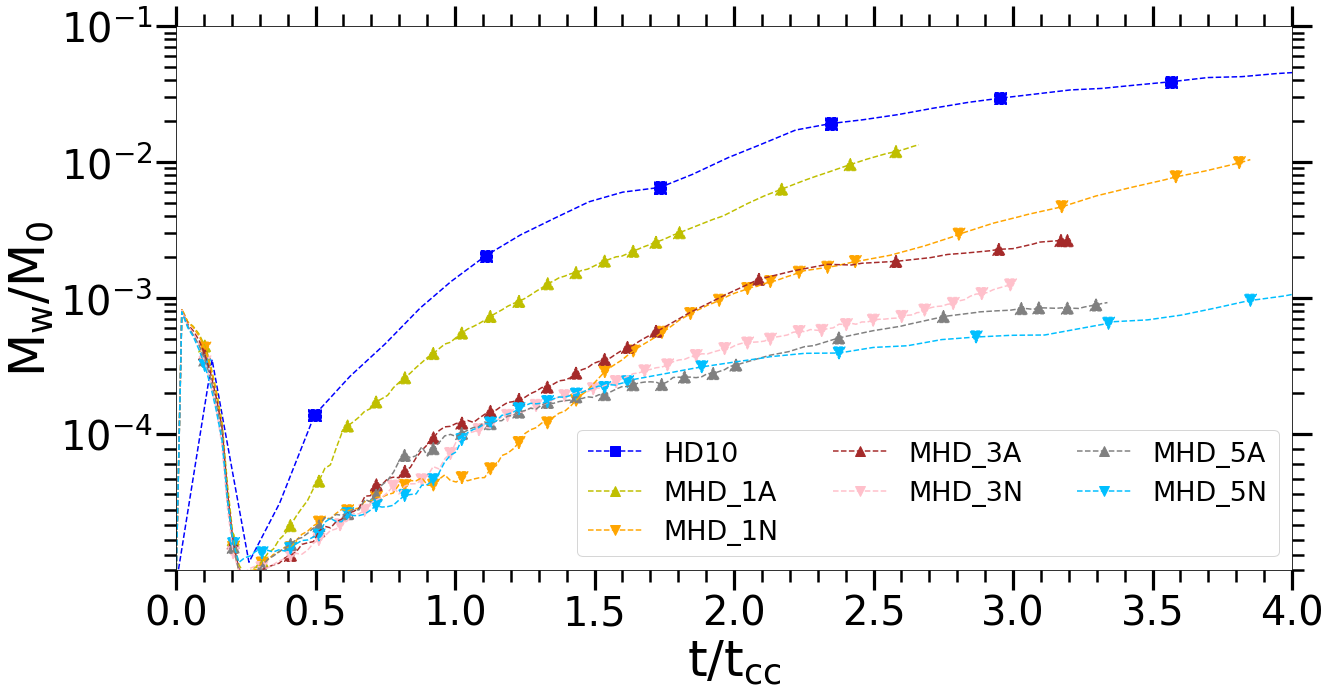}
    \caption{Time evolution of the normalised warm gas mass fraction for six magnetised variations of the \run{HD10} simulation. Magnetic fields reduce the amount of warm gas by up to two orders of magnitude over a few crushing timescales for the range of field strengths tested here. Reduction in the warm gas mass is more pronounced for the simulations where magnetic field lines are initially normal to the flow.}
    \label{plot:mhd_warm}
\end{figure}

As can be seen in Fig.~\ref{plot:MHD multislice}, magnetic fields have a significant impact on the morphology of cold clouds and their wake over time, but not enough impact to prevent the growth of the cold cloud mass (see Fig.~\ref{plot:mhd_cold}). In comparison to the reference simulation \run{HD10}, mass growth is slower for all magnetic field strengths tested here, and this suppression is greater for stronger magnetic fields irrespective of the initial field configuration. The warm gas mass fraction is similarly suppressed, as shown in Fig.~\ref{plot:mhd_warm}, with initially stronger magnetic fields leading to less warm gas mass. Interestingly, the trend in mass growth is not reversed between the warm and cold gas phases like it is for the simulations of clouds with different density (see Section~\ref{sec:Density Dependence}), where the clouds that grow fastest in cold gas mass tended to generate the least warm gas. 

\begin{figure}
    \centering
    \includegraphics[width=\columnwidth]{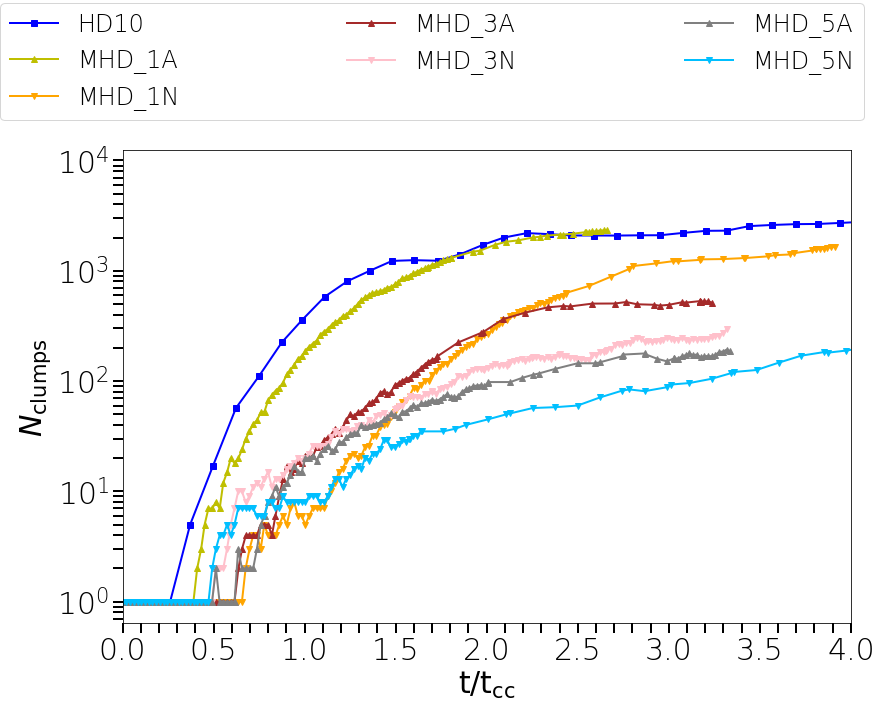}
    \caption{Time evolution of the number of clumps (top) for six simulations with different magnetic field orientations and magnetic field strengths of the \run{HD10} simulation. Higher magnetic field strengths, and fields initially orientated normal to the direction of flow, lead to less fragmentation.}
    \label{plot:mhd_nclumps}
\end{figure}

We also see a distinct difference in cloud evolution between the two magnetic field configurations tested here, with the perpendicular field suppressing mixing and allowing for more efficient cooling at all field strengths tested than the equivalent runs with initially aligned fields to the flow. For our highest field strength ($ 5 \rm \ \mu$G), there is a difference of approximately $8\%$ in the cold gas mass and a difference in the warm fraction of approximately $53\%$ at $3$ \tcc\ between the normal and aligned configurations. For a $1\ \mu$G field these differences are larger at 3 \tcc, at $110\%$ and $238\%$ in the cold and warm fractions, respectively. Under the assumption that mixing is driven by the Kelvin-Helmholtz instability, such as in the analytical solution by \citet{Chandrasekhar_1961}, this trend is surprising. If mixing is instability driven, a magnetic field with fields lines parallel to the fluid flow should stabilise the Kelvin-Helmholtz instabilities and slow the growth rate of unstable modes, with larger field strengths providing greater stabilising action. This is due to the magnetic tension force resisting the bending of fields lines. 

To explain the trends in the warm gas reported here, it is necessary to instead consider the morphology of the cloud and the topology of the field lines that result from the two initial field set-ups. The impact of magnetic fields on the warm gas in particular can be seen visually in Fig.~\ref{plot:MHD multislice}, which shows that at fixed field strength, the aligned configuration produces a significantly more extended wake than the normal configuration. This trend is confirmed in Fig.~\ref{plot:mhd_nclumps} which shows that at fixed field strength, there are fewer individual fragments downstream in the wake than for magnetic fields perpendicular to the flow. As discussed in Section~\ref{sec:Cooling Floor}, a more extended wake and greater fragmentation leads to a greater surface area of interface between the cold and hot gas, which increases mixing and production of the warm phase. This, in turn, drives increased condensation.

There are a several mechanisms suppressing the wake in the perpendicular field configuration. Firstly, there is a strong magnetic draping effect taking place near the head/s of the clump/s, such as that described by \citet{pfrommer2008} and more recently by \citet{SparrePfrommer2020}. As the clump moves through the hot \ac{icm}, it `sweeps up' magnetic field lines, which become pinched together near the head of the cloud. \citet{pfrommer2008} find that the magnetic tension associated with this draped layer can dominate over hydrodynamic drag forces in slowing the clump down and reducing fragmentation.
 Draping of the field does not happen as significantly in the case where the magnetic field is parallel to the flow velocity, since the cloud material can slip along with the field lines. In our case, magnetic draping is present at the head of the cloud, as can be seen by the increased field magnitude at the head of the clumps in the perpendicular field configuration in Fig.~\ref{plot:MHD multislice}. Perhaps the most important draping effect however is the suppression of Kelvin-Helmholtz instabilities when the field is tangled, which is not as pronounced if the field is uniform. This is due to a draping layer which suppresses Kelvin-Helmholtz instabilities \citep{SparrePfrommer2020}, which is more effective in a tangled-field scenario because in a uniformly magnetised wind the instabilities are suppressed only along the axis which is parallel to the field orientation, rather than along all axes.

\begin{figure}
    \centering
    \includegraphics[width=\columnwidth]{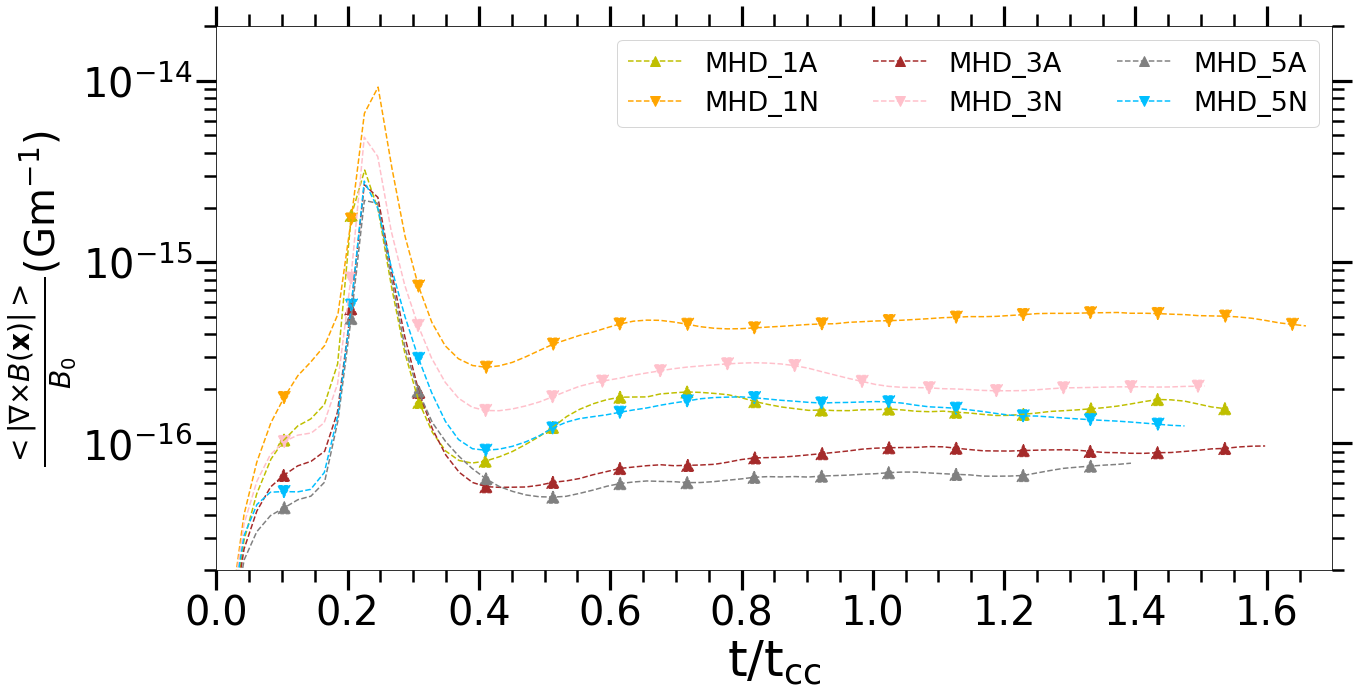}
 \caption{The average value of the curl of the magnetic field (mass-weighted) for all `non-hot' gas (i.e. all gas below the wind temperature). The perpendicular field configuration tends to have a larger value of the curl after the early cloud collapse than the parallel field equivalent.}
    \label{plot:curl_evolution_non_hot_gas}
\end{figure}

The second process acts on the wake behind the clump. As can be seen in the line-integral convolutions which trace the magnetic field direction in  Fig.~\ref{plot:MHD multislice}, the field morphology in the wake depends strongly on the initial magnetic field topology. For an initially parallel field, the magnetic field wraps around the cloud and wake, and a field configuration that is mostly parallel to the flow is maintained. In the perpendicular case, however, the field becomes very turbulent downstream of the wake. This tangling of the field lines creates magnetic tension, which prevents the wake from elongating as efficiently in the perpendicular field case, due to the Lorentz force on the plasma in this region. This effect is probably enhanced by the suppression of mixing instabilities by draping, with less clumpy material then being swept downstream in the wake.

To quantify this non-uniformity of the field in the wake and compare between the aligned and perpendicular field configurations, we plot the average curl of the magnetic field in non-wind ($<10^8$ K) fluid in Fig.~\ref{plot:curl_evolution_non_hot_gas}. We find that the perpendicular field always has a higher mass-weighted average of the magnitude of the $B$-field curl than the parallel field case in the mixed gas, by up to a factor of a few. This trend persists at later times as the value of the curl converges. 

\begin{figure}
    \centering
    \includegraphics[width=\columnwidth]{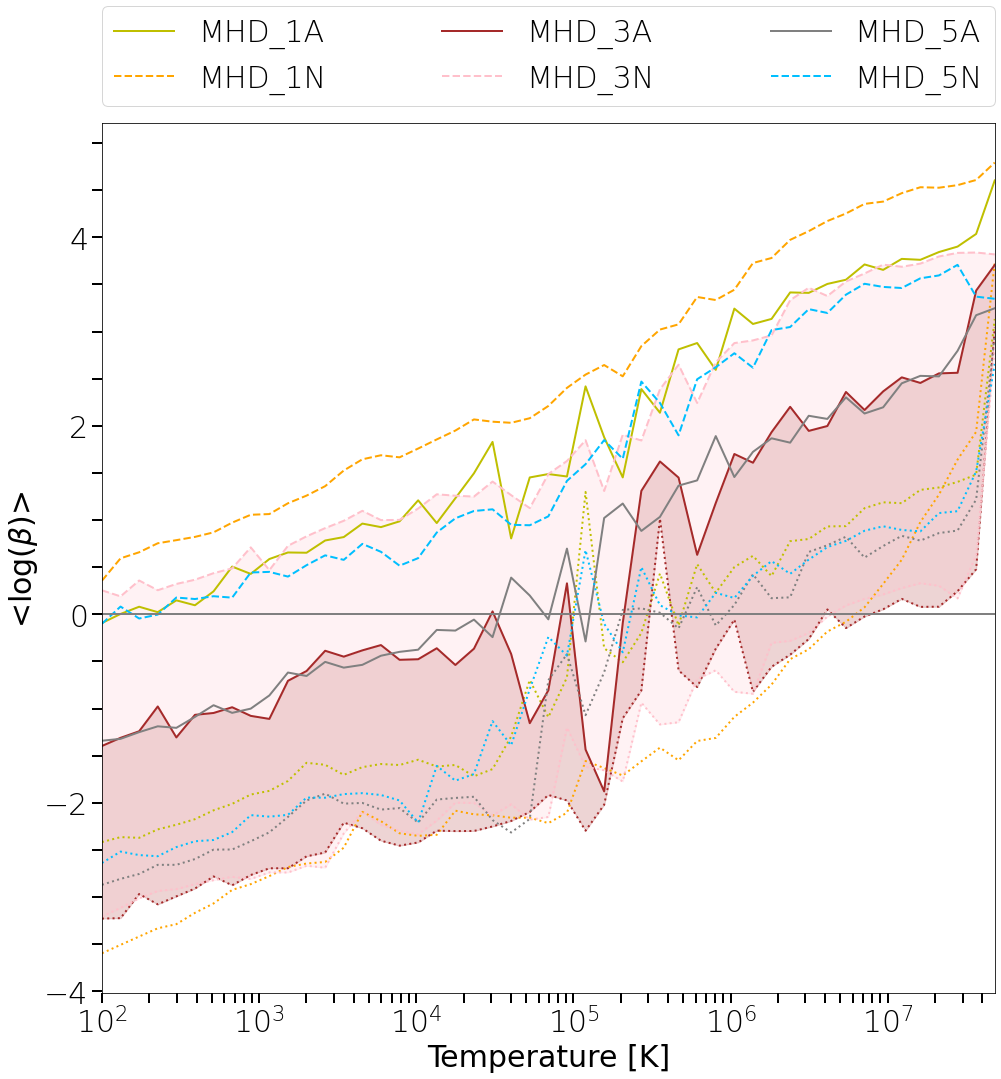}
    \caption{Distribution of plasma $\beta$ versus temperature for all magnetised simulations at $t=2.6$ \tcc. Solid (dashed) lines mark the upper $90$th percentile of the distribution at a given temperature for all runs with initially parallel (perpendicular) magnetic fields while dotted lines mark the lower $10$th percentile. The shaded regions highlight the range of values for \run{MHD\_3N} and \run{MHD\_3A}, respectively. Except in the range $10^5\,\mathrm{K} < T < 10^6\, \mathrm{K}$, lower limits for \run{MHD\_3N} and \run{MHD\_3A} are very similar. For runs with an initially perpendicular magnetic field, magnetic pressure support dominates for cold gas, while for initially aligned field the distribution is much wider.}
    \label{plot:magnetic_pressure}
\end{figure}

As discussed in Section~\ref{sec:KH vs Pressure Gradients}, mixing in our simulations is primarily driven by Kelvin-Helmholtz instability, but there is still some growth in the cold phase even when the cloud is approximately co-moving, which is likely due to the pressure-driven effects. With an initial plasma $\beta = 10^3 - 10^{4.5}$ ($\beta=2P_{\rm th}/B^2$, where $P_{\rm th}$ is the thermal pressure and $B^2/2$ is the magnetic pressure), magnetic pressure effects are negligible in the hot wind but $\beta$ naturally decreases as the gas cools until $\beta <1$ in the cold clumps (see Fig.~\ref{plot:magnetic_pressure}). The relative strength of magnetic pressure at a given temperature depends on the field configuration. Originally perpendicular fields produce a much wider range of $\beta$ values at a given temperature in comparison to aligned fields. However, for all simulations tested here, much of the warm gas remains on average dominated by thermal pressure ($<\log(\beta)> \gtrsim 0 $). The presence of magnetic fields, therefore, does not prevent the pressure-gradient-driven mixing of warm gas into cold gas, and at best merely slows the process down. Interestingly, the minimum $\beta$ as a function of temperature is the same for both simulations, but \run{MHD\_3N} has a much higher maximum  $\beta$ than gas of the same temperature in \run{MHD\_3A}. As a result, all cold clumps are magnetically dominated for \run{MHD\_3A}, while some cold clumps remain thermal pressure dominated for \run{MHD\_3N}.

\subsection{Thermal Conduction}
\label{sec:thermal_conduction}

In the presence of strong temperature gradients, such as those seen in and around cold clumps in the intracluster environment, free electrons transport heat parallel to the magnetic field lines \citep{Cowie1977}. This anisotropic thermal conduction is expected to smooth temperature, pressure and density gradients in multiphase systems \citep{Vieser2007}. For sufficiently high field strengths, it will also suppress orthogonal instabilities \citep[e.g.][]{Parrish2012} as momentum and conductive heat transfer can be restricted almost exclusively to paths along field lines. As well as stabilising instabilities and thus reducing the proportion of the gas in the catastrophically-cooling warm phase, conduction has the potential to directly slow down the cloud growth by offsetting radiative cooling losses with energy transferred from the surrounding hot medium. 

Anisotropic conduction is computationally expensive to model in simulations and for this reason has frequently been omitted in previous work on the cloud crushing problem, or approximated using analytic assumptions based on the morphology of the magnetic field, which translates to a numerical ``fudge factor'' in the classical Spitzer formula \citep{Spitzer1962}. In this paper, we explicitly model the conduction within \ramses as part of our full MHD set-up. All simulations with conduction are detailed in Table~\ref{tab:mhd runs}, and include runs with either isotropic (post-fixed with $X\_{\rm i}$) and anisotropic conduction ($X\_{\rm a}$). For each type of conduction we test a strong, an intermediate, and a weak field in both the perpendicular and aligned configuration. We shall continue referring to the magnetised runs without conduction as `MHD' runs.

\begin{figure}
	\includegraphics[width=\columnwidth ]{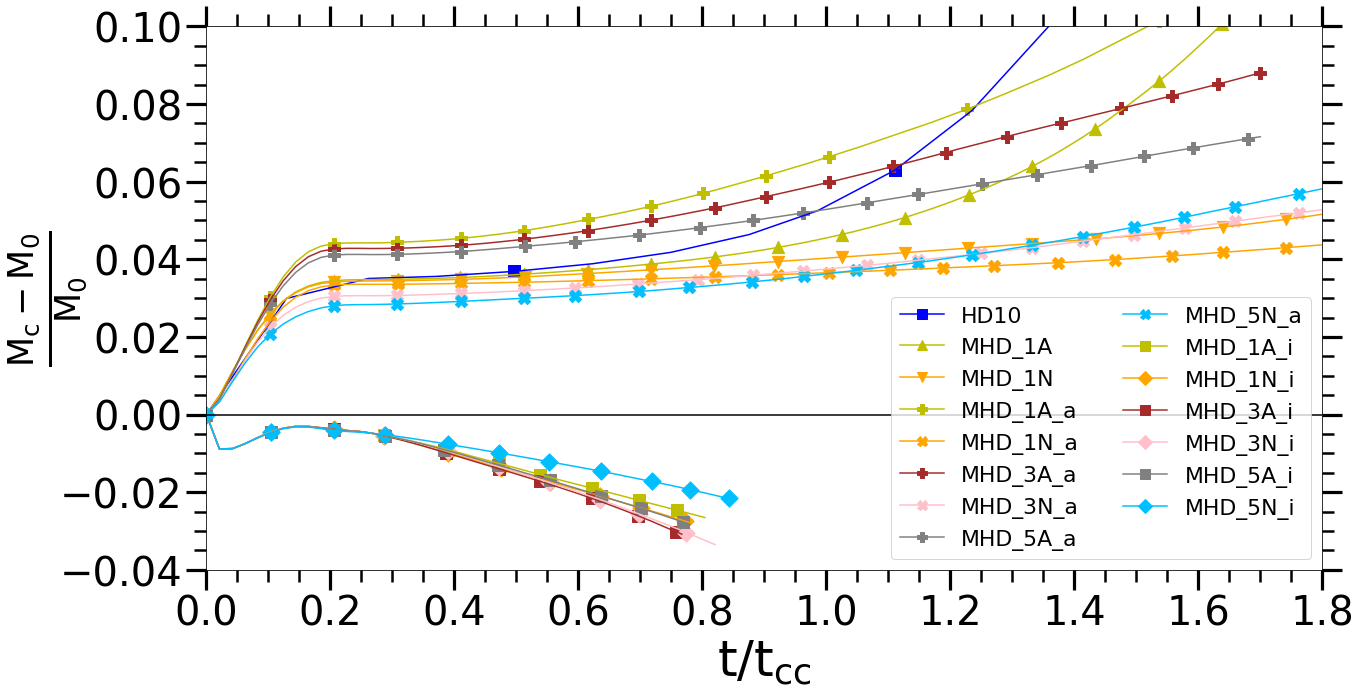}
    \caption{Time evolution of the normalised cold fraction for cold clouds simulated with magnetic fields and thermal conduction. Isotropic thermal conduction causes clouds to evaporate in the hot wind, while anisotropic thermal conduction is insufficient to prevent their growth. Note that the growth rate for MHD runs is slow within the first \tcc, hence the linear scale on the vertical axis.}
    \label{plot:conduction_cold}
\end{figure}

\begin{figure}
    \centering
	\includegraphics[width=\columnwidth ]{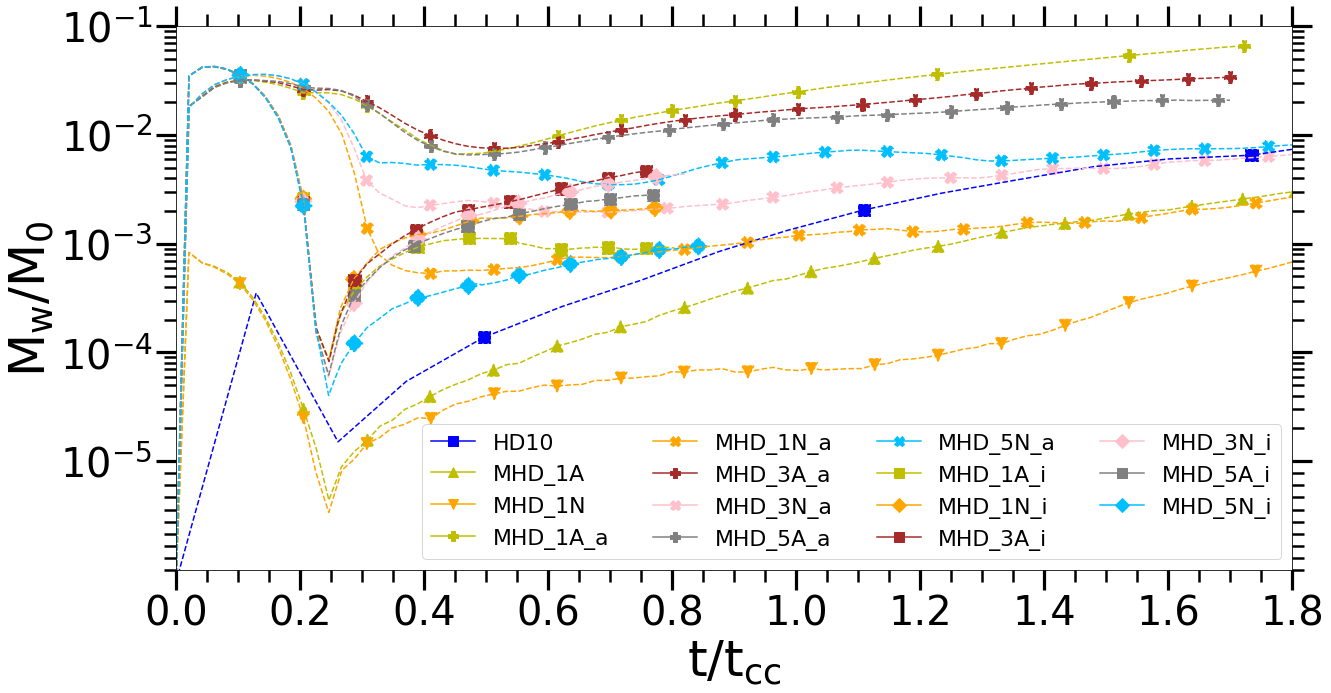}
    \caption{The normalised warm mass fraction for all simulations with magnetic fields and thermal conduction. Both isotropic and anisotropic thermal conduction increase the amount of warm gas in comparison to simulations without conduction. This effect is generally stronger for anisotropic than isotropic thermal conduction.}
    \label{plot:conduction_warm}
\end{figure}

\begin{figure*}
	\includegraphics[width=16cm]{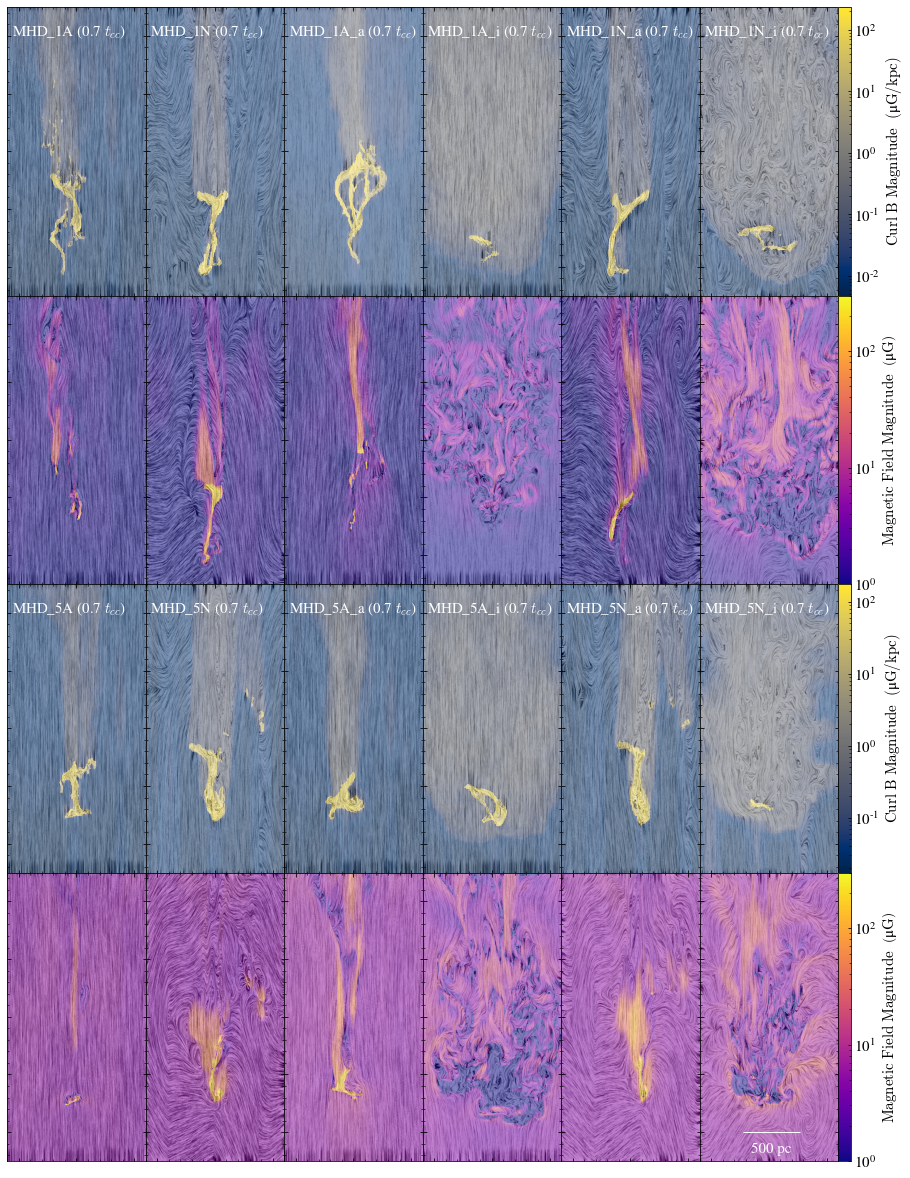}
    \caption{The density-weighted magnetic curl projection (top panel) and magnetic field magnitude slice (bottom panel) for the eight MHD runs with thermal conduction, as well as four corresponding non-conduction MHD runs. The orientation of the magnetic field lines is shown via the line integral convolution.}
    \label{plot:Conductionmultislice}
\end{figure*}

The addition of thermal conduction influences both the amount of cold and warm gas, with isotropic conduction producing significantly wider wakes than anisotropic conduction (see Fig.~\ref{plot:Conductionmultislice}). While the remaining cloud cloud stays very compact in the isotropic conduction phase, any material ablated is mixed into a wide, turbulent wake which is quickly heated to hot gas temperatures. By contrast, in the anisotropic case, wakes remain much narrower and the ablation of diffuse material from the back of cloud produces the warm gas seen here.

The clearest impact of conduction is seen in the cold gas, where cloud evolution separates into two very distinct modes depending on whether the conduction is isotropic or not (see Fig.~\ref{plot:conduction_cold}): isotropic conduction causes the cold gas mass to decrease over time, similarly to but more slowly than what is observed in the non-cooling runs shown in Fig.~\ref{plot:coolvsnocool_cold_mass}. For all runs with isotropic conduction, the hot \ac{icm} can supply sufficient thermal energy to offset cooling and prevent the formation of new cold gas from the mixed phase. The peak in the condensed mass just after $0.15$ \tcc\ (where the cold mass losses are briefly almost nullified) is likely due to the high fraction of warm gas formed within $0.1$ \tcc, which may manage to condense some cold gas before the warm fraction drops after $0.2$ \tcc. From there on, the cold mass is reduced by around $1\%$ by $0.4$ \tcc \ and by $2-3\%$ by $0.8$ \tcc. This mass loss rate is around 10 times slower than the non-cooling run of the same density, \run{HD10\_nc} whose cold mass is reduced by $30\%$ by $0.9$ \tcc. We conclude that isotropic thermal conduction is able to offset radiative cooling in and around cold clouds in galaxy clusters and to prevent their growth on long timescales. 

As expected, conduction acts to generally increase the warm gas fraction compared to the MHD runs, due to its ability to transfer heat from the background hot wind to the mixed phase. This is shown in Fig.~\ref{plot:conduction_warm}. This warm gas can no longer be classified as a pure "mixed" phase, as heat is now allowed to flow along temperature gradients, i.e. warm gas is now also generated through the heat flux from the hot \ac{icm} into the cold cloud fluid. As a result, more warm gas is created early on, up to $\sim 100$ times the amount seen in simulations without conduction at $0.1$ \tcc, since conduction timescales are shorter than the dynamical mixing timescales. The conduction runs produce more warm gas than any of the purely hydrodynamical or MHD runs. For example \run{MHD\_1N\_a} and \run{MHD\_1N\_i} produce $13$ and $37$ times as much warm gas, respectively, as \run{MHD\_1N} at 0.6 \tcc. 

\begin{figure}
    \centering
	\includegraphics[width=\columnwidth ]{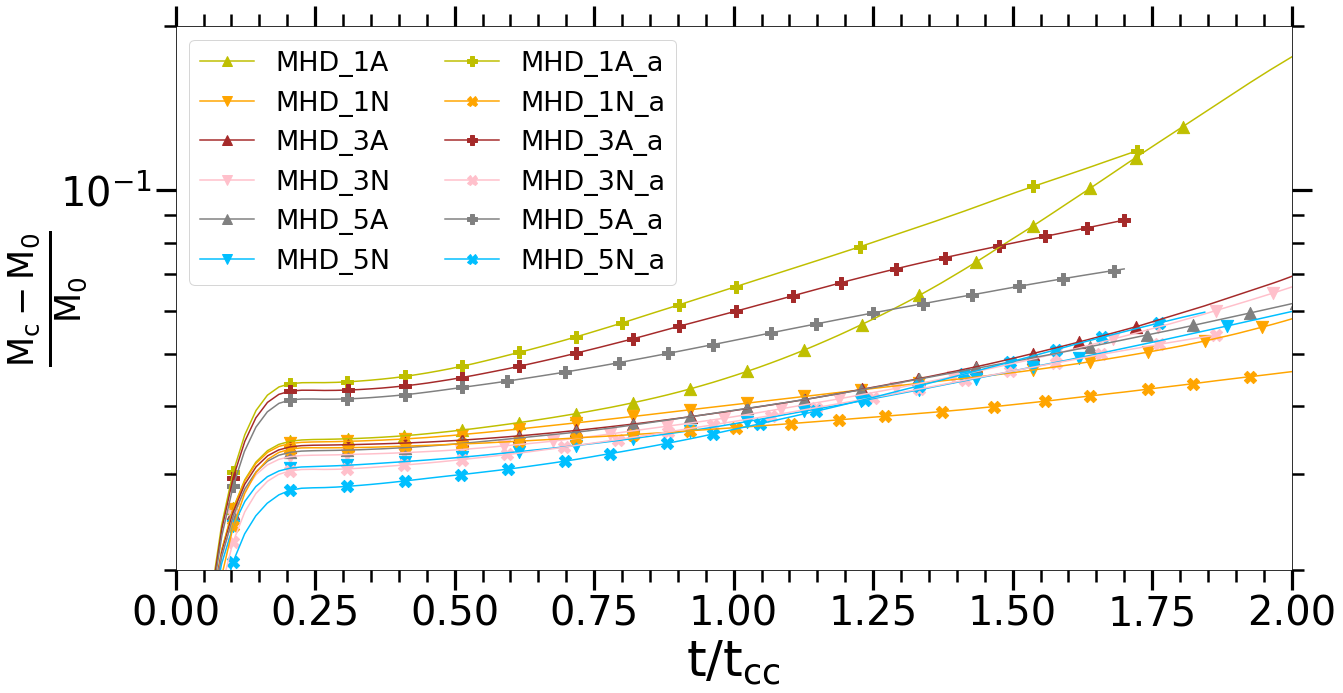}
    \caption{Time evolution of the normalised cold fraction for all cold clouds that grow in Fig.~\ref{plot:conduction_cold}. Anisotropic conduction makes little difference to the evolution of the cloud for a magnetic field that is normal to the direction of flow, but boosts the growth of cold gas for an initially aligned magnetic field.} 
    \label{plot:conduction_cold_growing_runs}
\end{figure}

For all anisotropic cases we see evolution more in line with the other simulations presented in this paper: clouds grow in mass over time, as can be seen in more detail in Fig.~\ref{plot:conduction_cold_growing_runs}. Apart from an early boost in cold gas mass for the cases with an aligned magnetic field, the growth rate of cold gas in the presence of anisotropic conduction is reduced in comparison to the equivalent non-conducting runs. The impact of anisotropic thermal conduction on both the cold and the warm phase can be understood in the context of the length-scales of conduction in and around cold clouds. Following Section 3.1 of~\citet{Li2020}, the heat precursor in the hot phase is $\lambda_e\simeq 100\,\rm pc$. The skin depth of heat conduction in the cloud is $\lambda_{\rm skin}=\lambda_e /\chi\simeq 1\,\rm pc$. As a result, the thermal conduction has no direct effect on the cold cloud (such as suppressing KH instabilities or reheating cold gas to the warm phase), but it warms up the mixing layer on a $100$~pc scale. This leads to the increased amount of warm gas observed here, but also explains why this warm gas cools more slowly into cold gas. Note that the effect of thermal conduction is directly proportional to density of the cloud. For clouds with $\chi<100$  the skin depth will quickly increase, and thermal conduction should have a very significant effect on the cloud structure and the number of cloudlets.

\begin{figure}
    \centering
    \includegraphics[width=\columnwidth]{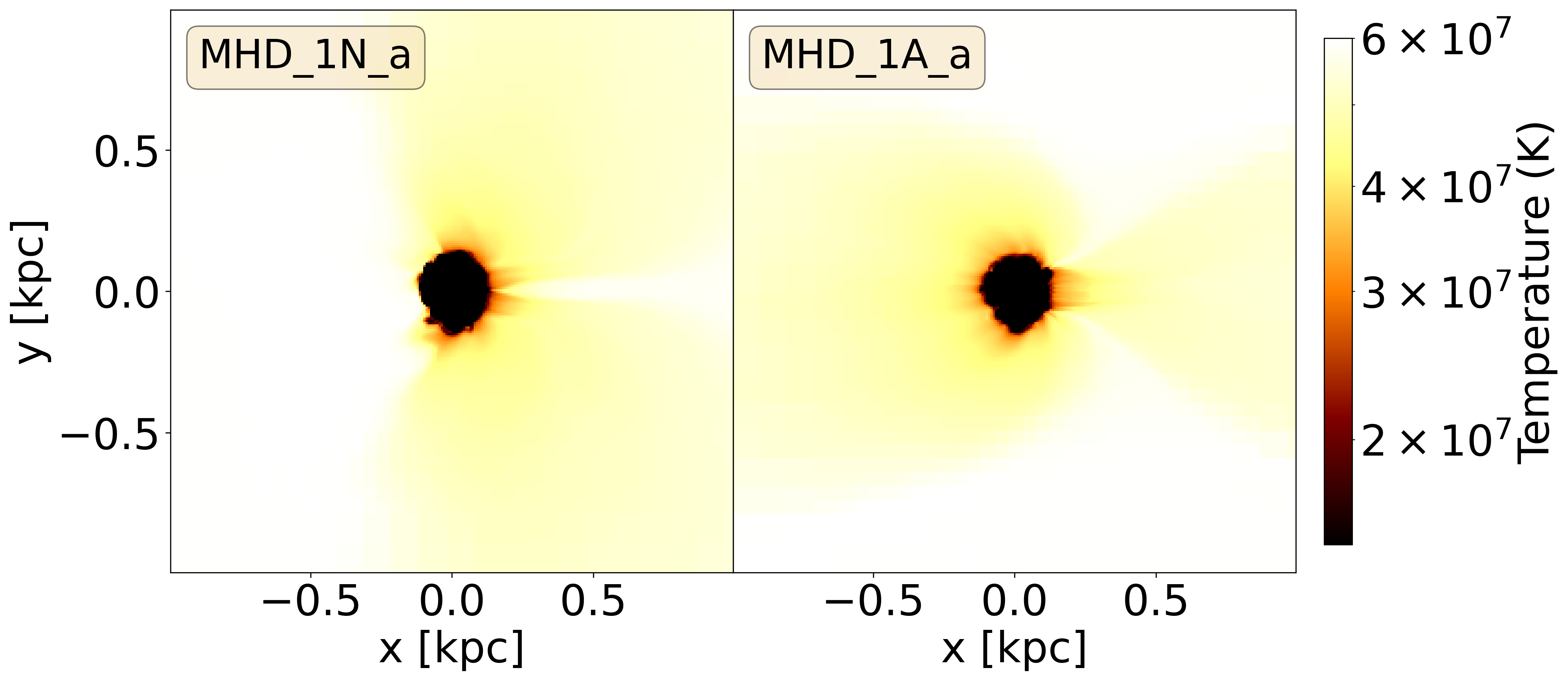}
    \caption{Slice plot of the temperature in \run{MHD\_1N\_a} (left) and \run{MHD\_1A\_a} (right) at $t=0.18 \, t_{\rm cc}$. Limits were chosen to highlight the structure of the plume pre-cooled by thermal conduction.}
    \label{plot:plume}
\end{figure}

To understand the early boost in cold gas mass for aligned magnetic fields in the presence of anisotropic thermal conduction (compare \run{MHD\_1A\_a} to \run{MHD\_1A} and \run{MHD\_1N\_a} in Fig.~\ref{plot:conduction_cold_growing_runs}), one needs to look in detail at where anisotropic conduction extracts the energy from the hot wind. As can be seen visually in Fig.~\ref{plot:plume}, in the aligned case (\run{MHD\_1A\_a}), thermal energy is extracted upstream of the cloud as it flows down the magnetic field lines. This creates a cooler plume of gas ahead of the cloud, which, due to the loss of thermal pressure, also becomes somewhat denser. Due to the flow direction, this gas then impact the mixing layer of the cloud, where the early cooling at $t < 0.2 \, t_{\rm cc}$ occurs. With a perpendicular magnetic field (\run{MHD\_1N\_a} in Fig.~\ref{plot:plume}), a similar plume forms next to the cloud but due to its location, the plume in this case flows downstream without interacting further with the cloud. For this reason, the cloud in the aligned case is fed by pre-cooled, denser gas early on which reduces the cooling times in the mixing layer. By contrast, the cloud in a perpendicular field configuration is fed by pristine wind gas, like in the non-conducting case. As a result, clouds in an aligned field configuration grow faster. We expect this early boost to cold gas mass to be almost entirely restricted to magnetic fields parallel to the flow. In most other configurations, magnetic field lines will quickly drape over the cloud \citep[see][for a careful study of draping]{SparrePfrommer2020}, suppressing conduction at the surface and maintaining large temperature gradients with reduced `mixing'. Heat transport would also be restricted to chaotic paths in the tangled field, which may further reduce the efficiency of heat transfer and the mass fraction of the warm phase.

From our simulations with thermal conduction, it is clear that conduction adds significant complexity to the problem of cloud growth; on one hand, heat conduction from the wind to the boundary layer increases the warm fraction, which, in turn, can increase the growth depending on where the pre-cooled gas is located. Conversely, the same conduction channel can supply heat from the wind to this warm fraction (whether generated by conduction or Kelvin-Helmholtz mixing), and help to offset radiative losses, increasing the cooling timescale and consequently decreasing the condensation rate. However, we find that with anisotropic thermal conduction, and a range of different magnetic field morphologies, cold clouds in the hot ICM continue to grow, albeit more slowly than in the non-magnetic case. 

We do note that the magnetic field and conduction runs are very computationally expensive, and as such we are only able to push them to a few crushing times at most. This may mean that we miss important late-stage evolution while the cloud is entrained, which could be different to the early stage evolution where the cloud is mainly being mixed and accelerated and generates a 'tail' \citep[see e.g.][]{ GronnowTepper-Garcia2018, Gronke2019, Kanjilal2021}. This entrainment phase usually starts later for clouds in magnetic fields, and begins after multiple \tcc, possibly beyond the time we are able to simulate for for our clouds. We therefore advise that while our results apply to the initial stages of cloud growth, without simulations able to follow the late-time evolution of magnetic field and conduction runs one should be cautious to apply the same findings to the latter stages of cloud evolution.

\section{Discussion and Conclusions}
\label{sec:conclusions}

\begin{figure}
\centering
	\includegraphics[width = \columnwidth]{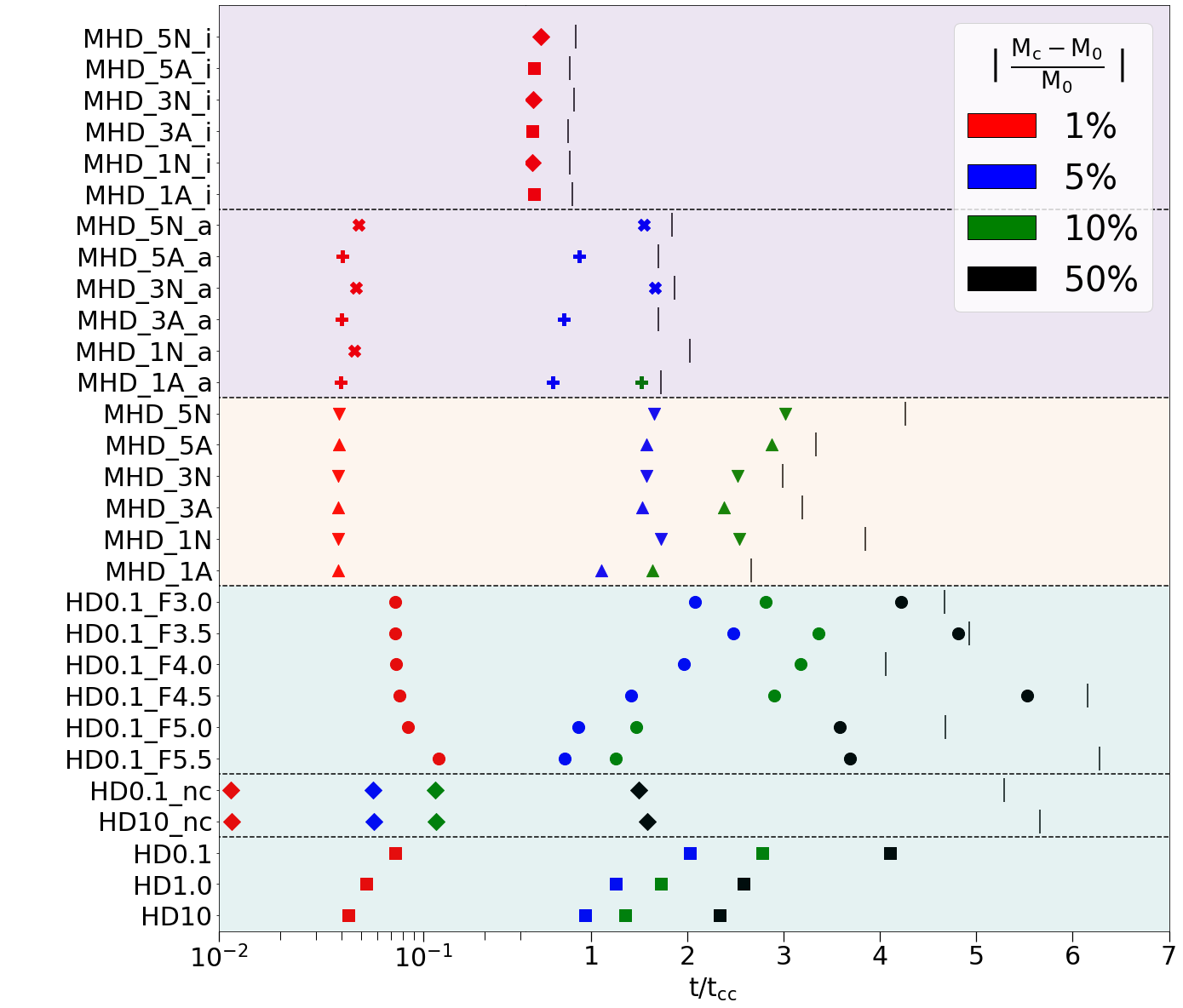}
    \caption{Time it takes for clouds to increase (or decrease, in the case of no-cooling or isotropic conduction) their cold gas mass by $1\%$, $5\%$, $10\%$, and $50\%$. Where necessary, the time was linearly interpolated between existing simulation outputs. All MHD simulations are variants of \run{HD10}. General trends that have been described in detail in previous sections stemming from the effects of magnetic fields, various cloud densities, and differing temperature floors can be easily seen. The $x$-axis is logarithmic up to $t/$\tcc $=10^{-0.5}$ and linear thereafter. The vertical black markers show the final time reached for each run.}
    \label{plot:cold_percentage}
\end{figure}
\begin{figure}
    \centering
	\includegraphics[width = \columnwidth]{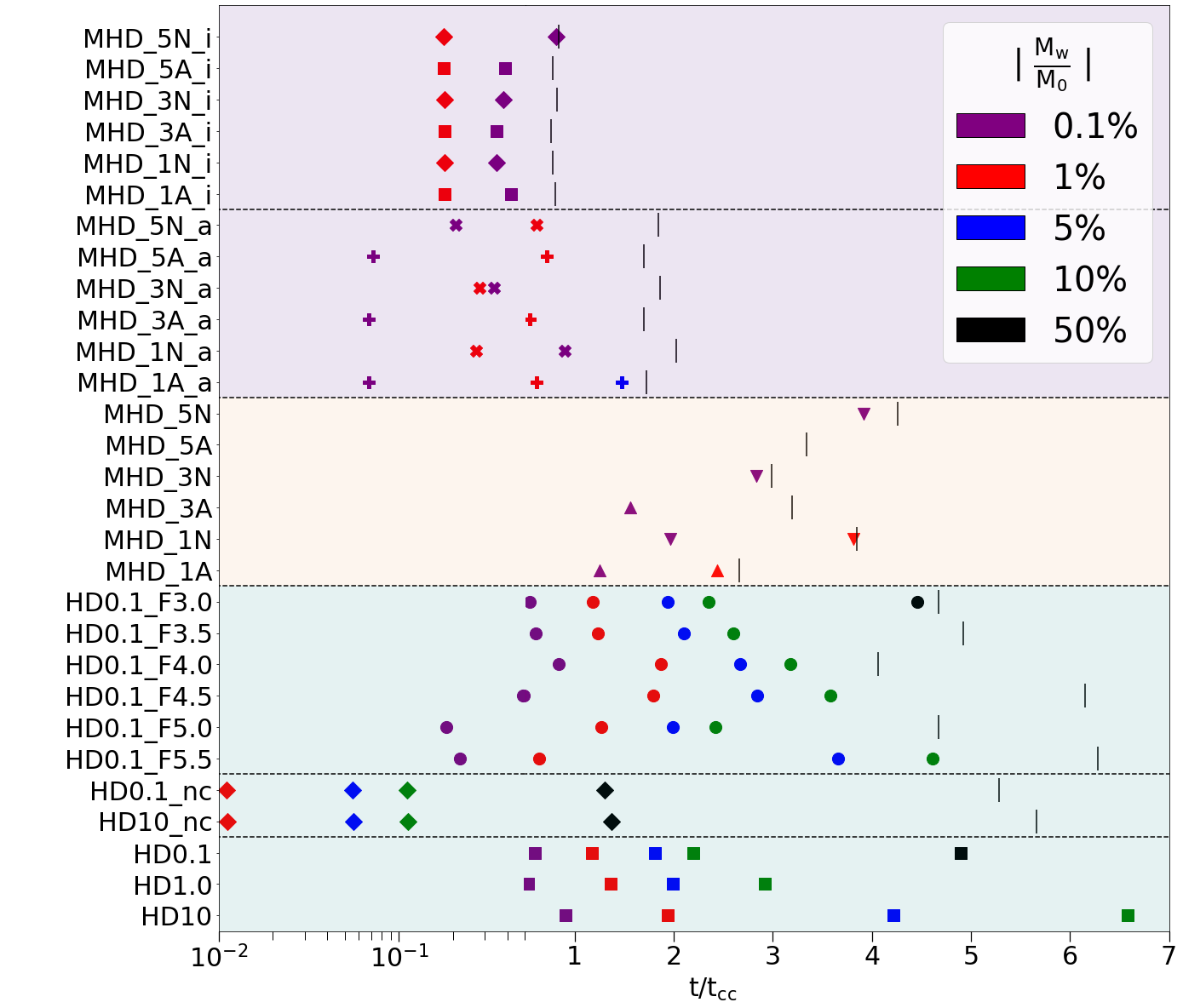}
    \caption{Time it takes for clouds to increase (or decrease, in the case of no-cooling or isotropic conduction) their warm gas mass by $1\%$, $5\%$, $10\%$, and $50\%$. Where necessary, the time was linearly interpolated between existing simulation outputs. The bottom axis is logarithmic up to $t/$\tcc $=10^{-0.5}$ and linear thereafter. The vertical black markers show the final time reached for each run. }
    \label{plot:warm_percentage}
\end{figure}

In this work we have performed high-resolution simulations of $500$~pc-scale cold clouds in a hot \ac{icm} environment. We have gradually introduced an increasingly more complex physical picture, by including radiative cooling, small-scale reheating, magnetic fields with different topologies and strenghts, and (anisotropic) thermal conduction, and investigated the impact of each individually, with the aim of better understanding the evolution of cold gas within a cluster environment. 

Using a set of cloud parameters typical of fragmented filaments in the realistic hot \ac{icm}, we confirm that radiative cooling is the dominant physical process that determines the continued evolution of clouds on tens of Myr timescales. Except for those simulations with deliberately unphysical models (no cooling or isotropic thermal conduction), our cold cloud grows in mass for all parameters probed here. However, as can be seen in Fig.~\ref{plot:cold_percentage}, how fast the cold cloud mass grows is strongly a function of the detailed physics being modelled. The same is true for the warm gas mass, as summarized in Fig.~\ref{plot:warm_percentage}. Specifically we find that
\begin{itemize}
\item The presence of radiative cooling breaks the scale-free nature
of the cloud crushing problem, by introducing an additional characteristic timescale which depends on cloud density, temperature and metallicity.
\item Due to radiative cooling in the warm phase, cold clouds can significantly increase their mass on cloud-crushing timescales, as the dynamical generation of the warm phase via Kelvin-Helmholtz instabilities is the limiting factor for growth. We find a drop-off in cloud growth as the cloud starts to become co-moving with the wind, as mixing is driven mainly by Kelvin-Helmholtz instabilities rather than pressure gradients induced by cooling.
\item As expected, initially denser clouds show faster mass growth due to the more rapid condensation of gas from the warm to the cold phase. Lower density clouds form lower density fragments post-shattering, and this results in greater mixing of cold gas into a warm phase. 
\item The exact minimum temperature to which the cloud can cool (the temperature floor) splits the cloud evolution into two modes, which we link to theoretical predictions of post-shattering clump lengthscales: above a certain temperature ($\sim 10^4$~K) the cloud ablates in the wind, with clumplets being formed dynamically and a significant cloud ``head'' retained. Instead for low temperature cooling, the cloud shatters due to cooling-induced pressure instability, forming many fragments, and takes more time to start generating warm gas. We therefore conclude that care must be taken to model low temperature cloud cooling, that should be ideally tuned to observations and/or well-motivated models of sub-cloud heating. 
 \item Adding magnetic fields has little impact on the early evolution of the cold cloud mass but significantly reduces the amount of warm gas in the wake. This is due to the fact that magnetic fields suppress the mixing of gas in the wake of the cloud which occurs even for the weakest $B$-fields explored here ($ 1 \  \mu$G). Higher $B$-field strengths are most efficient at suppressing cold cloud mass growth as well as the amount of warm phase. 

 \item The combined effects of magnetic pressure and magnetic tension act to keep the hot and cold/warm phases separate at the boundary layer and reduce the mixing efficiency of the Kelvin-Helmholtz instabilities. Interestingly, the suppression of cloud growth is most pronounced for the field line configuration that is initially perpendicular to the wind. This stems from the fact that these $B$-fields drape around the head of the cloud, increasing their stabilizing effect in the surface layer. Also, the wake region becomes highly disordered, which reduces the effective surface area of the cloud relative to the case where the $B$-fields are initially aligned with the flow. 

\item If isotropic, thermal conduction can deposit sufficient heat from the hot \ac{icm} into the cold and warm phases to prevent gains in the cold mass. In this case, the cloud evolution follows a destruction mode evolution which more closely resembles the traditional no-cooling cloud crushing solution, although at a rate around an order of magnitude slower.
\item If the conduction is anisotropic the cloud instead grows similarly to the non-conduction hydrodynamic and MHD runs because the the efficiency of heat transport via electrons is greatly reduced. The \ac{icm} is then unable to efficiently heat the cold cloud. Thermal losses via radiative cooling in the warm gas dominate again, albeit on a slower timescale than without conduction. 
\item This has important implications for both the \ac{icm} and other environments: simulations which either do not include magnetic fields or else do not explicitly model the conductive heat flux will not accurately account for changes to the heating efficiency caused by the specific orientation of the magnetic field lines. The presence of a magnetic field in any orientation, even if at the lower end of galaxy cluster values, will prevent conduction from destroying cold clouds.
\item Anisotropic thermal conduction can give early cold gas production a boost, if the magnetic field is aligned with the flow. Otherwise, the cold gas mass evolution behaves very similarly to the non-conducting case. In both cases, the amount of warm gas is significantly increased even in comparison to the non-conducting, and even the non-magnetic case, as warm gas is created both through mixing and through hot gas that looses thermal energy though conduction.
\end{itemize}

In general, we find that both modes of evolution (whether ablation from the surface of the cloud, or shattering of the original cloud into smaller cloudlets) impact the long-term evolution of the total cold gas mass. Another important driver, not surprisingly, is the total amount of warm gas produced in and around the cloud. This, in turn, depends on the mixing efficiency in the wake of the cloud, as well as on the efficiency of reheating such mixed gas via thermal conduction. We conclude that adding magnetic fields and anisotropic thermal conduction does not change the overall behaviour of cold clumps, but the relevant timescales do change, as well as the morphology of the mixed wake.

The growth of the clouds and the further fragmentation of the original parent clump may have important implications for the nature of the cold clump accretion \citet{Gaspari2013} onto the central \ac{bcg}. The shattering and/or ablation of cold clouds, as well as the growth in the cold mass, could alter the evolution of the cluster as a whole. Our results are also an important consideration when attempting to explain the so-called cooling flow problem, whereby feedback processes from the central \ac{agn} in the galaxy cluster are theorised to heat cold gas to observed levels, and prevent cooling of hot gas. We find that the cold gas mass in our simulations can grow by a factor 10 within a few $100$ Myr (although other cluster processes that we have not explicitly modelled would be expected to also affect cloud evolution on these timescales). Any proposed heating mechanism within clusters should be able to explain the observed cooling times and gas temperatures whilst also taking this extra cold mass into account. 

We note that the cluster parameter space is very large, and so it is difficult in practice to build a wholly comprehensive picture of cloud evolution in these systems through simulations that cover the entire range of free parameters. Instead, we have presented detailed studies for a typical fragment as those seen in the cluster-scale hydrodynamic simulations of \citet{Beckmann2019} and the magnetised version of the same cluster discussed in \citet{Beckmann2022,Beckmann2022b}. For our simulated cloud, we have found robust general trends governing cloud evolution under increasingly complex physics, such as magnetic fields and thermal conduction, and have shown that even with magnetic fields and thermal conduction, the continued evolution of individual cold clouds cannot be ignored when studying the cooling flow problem of galaxy clusters.

\section*{Acknowledgements}

FJ developed code, ran simulations, analysed data, interpreted results and wrote the manuscript. RSB initiated the project, developed code, analysed data, interpreted results and wrote the manuscript. DS and YD interpreted results, provided discussion and edited the manuscript. We would like to thank the referee, Prof. Peng Oh, for a constructive referee report that improved the paper. FJ and RSB thank Newnham College, Cambridge, for financial support. This work was supported by the ERC Starting Grant 638707 ``Black holes and their host galaxies: co-evolution across cosmic time" and by STFC. This work was performed using resources provided by the Cambridge Service for Data Driven Discovery (CSD3) operated by the University of Cambridge Research Computing Service (www.csd3.cam.ac.uk), provided by Dell EMC and Intel using Tier-2 funding from the Engineering and Physical Sciences Research Council (capital grant EP/P020259/1), and DiRAC funding from the Science and Technology Facilities Council (www.dirac.ac.uk). DiRAC is part of the National e-Infrastructure.

\section*{Data Availability}

The data used in this work may be shared on reasonable request to the corresponding author.



\bibliographystyle{mnras}
\bibliography{MAIN} 




\appendix
\section{Relative Contribution of Shattering vs Wind} \label{appendix: shattering_vs_wind}    
 \begin{figure}
	\includegraphics[width=\columnwidth ]{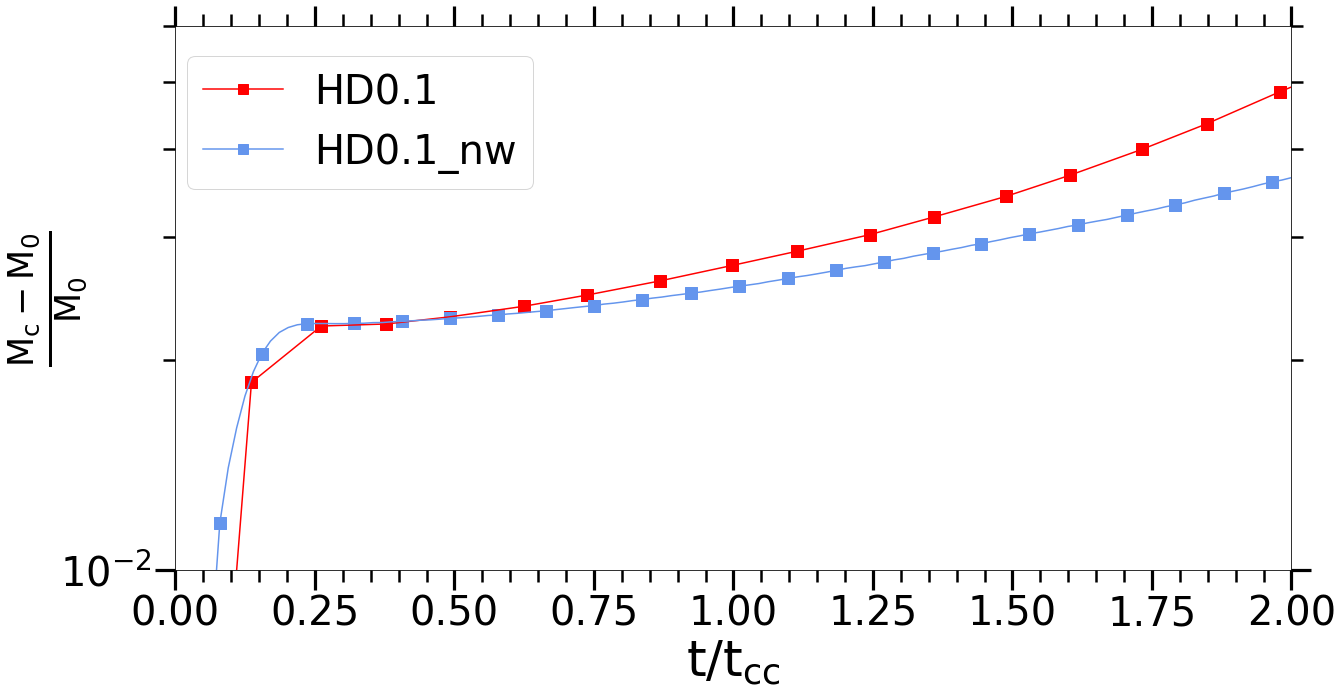}
    \caption{Here we plot our fiducial \run{HD0.1} against a no-wind version \run{HD0.1\_nw}, for which the only difference is that the wind speed is set to $0$ km/s, such that the run models a clump that is initially stationary within the \ac{icm}. No significant difference in the cold gas mass growth is seen in the shattering-dominated timescale $\lesssim 0.5$ \tcc. Beyond this the wind contribution to mixing starts to dominate and the no-wind run lags behind in mass growth.} 
    \label{plot:shatter vs wind}   
\end{figure}    
To investigate the relative contributions to mixing and subsequent cold mass growth from shattering versus wind-mixing, we run a version of our flagship cloud \run{HD0.1} with the cloud stationary, in the no-wind run \run{HD0.1\_nw}. As expected the mass growth rate post-shattering is reduced in the absence of a wind, as the degree of mixing via fluid instabilities is reduced. The mass growth rate during shattering \textit{is} similar, and this gives us a measure of the shattering timescale where cloud collapse and shattering dominates over the contribution from the wind. From Fig \ref{plot:shatter vs wind} we see that this shattering timescale is approximately $0.5$ \tcc.

\section{Convergence} \label{appendix: convergence}
\begin{figure}
	\includegraphics[width=\columnwidth ]{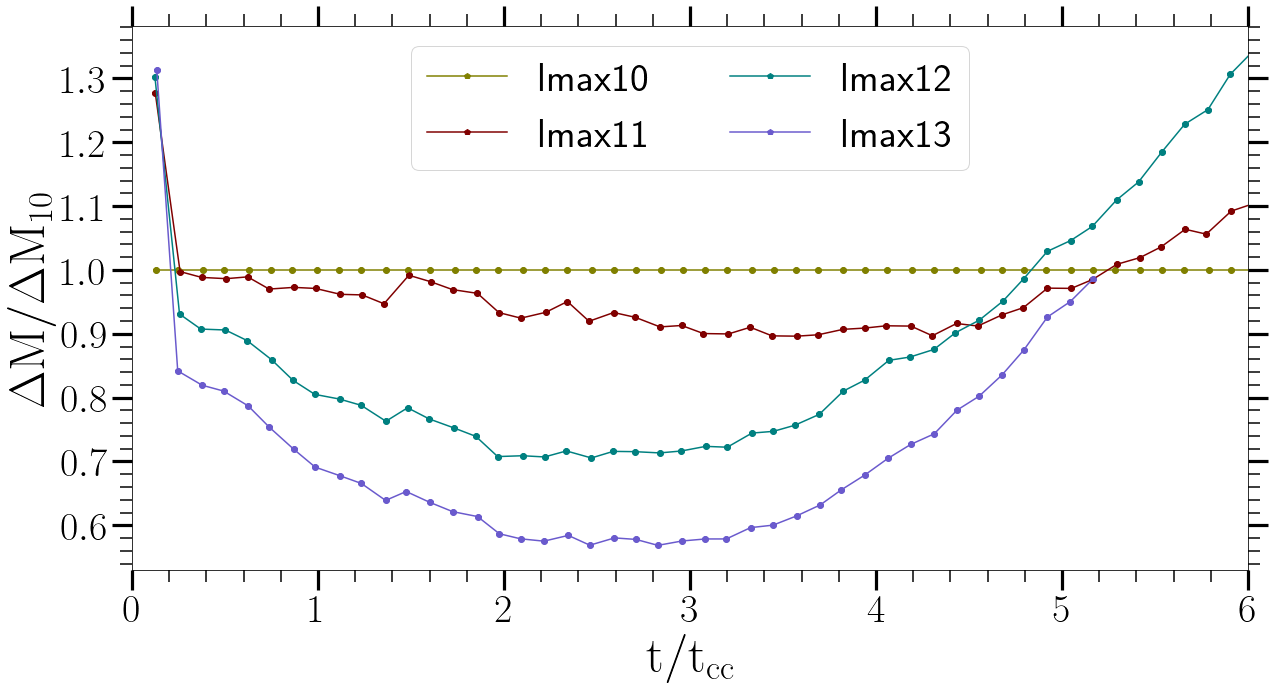}
    \caption{The change in cold mass against \tcc \ for a range of different resolutions for the same simulation run \run{MHD\_1N}, with respect to the minimum resolution run which has {\sc level\_max} equal to 10. While there is an initial paucity of old gas at early times with a higher resolution, this is reversed at around 5\tcc. This indicates that if the resolution were to be increased further, the growth effect we observe should be amplified even more.} 
    \label{plot:convergence plot}
\end{figure}    

In this paper we test a significant range of magnetic field strengths. The computational expense becomes quite large for high-strength fields, where the Alfv\'{e}n velocity becomes large and the time-step correspondingly small, and also when conduction is modelled. A sensible resolution must therefore be set to make our production runs feasible. Here we would like to test the convergence of our key results when adopting different resolutions for one representative run. The ``resolution parameter'' that we use in \ramses is the maximum refinement level {\sc level\_max}, which gives the number of refinements we allow the grid to make. In a maximally refined grid there will be $8^{\Delta x-1}$ leaf (terminal) cells, where $\Delta x\equiv$ {\sc level\_max}. We test the same simulation \run{MHD\_1N} (see Section~\ref{sec:mhd}) with a range of values for {\sc level\_max} in order to determine the convergence behaviour of our simulations, whilst keeping the minimum refinement level {\sc level\_min}, which sets the number of cells in the coarse grid, constant. The runs detailed here have a box size of ($195$~kpc)$^3$, so the maximum spacial resolution is $195/2^{\Delta x-1}$. This is the same as for all non-MHD runs performed in this paper. The MHD and conduction runs have a box-size ($303$~kpc)$^3$ (and also used a {\sc level\_min} value of $6$ as opposed to the value of $5$ used for the hydro runs). We plot the change in cold mass for a range of different maximal resolutions below with respect to the lowest resolution simulation. We find that for times less than 5\tcc there is less cold gas when the resolution of the simulation is increased, however the trend is reversed later on and we expect that our lower resolution simulations are giving us a lower bound on the long-term growth of the cold phase.


\bsp	
\label{lastpage}
\end{document}